\documentclass[apj,twocolumn,twocolappendix,numberedappendix]{openjournal_deluxetable_v2}
\usepackage{newtxtext,newtxmath,enumitem,multirow}
\usepackage{setspace}
\usepackage[utf8]{inputenc}
\usepackage[T1]{fontenc}
\usepackage[breaklinks,colorlinks,allcolors=blue,citecolor=blue,urlcolor=blue]{hyperref}
\usepackage{orcidlink}
\pdfoutput=1 
\newcommand{\mstar}{{\mbox{$M_{\rm star}$}}}
\newcommand{\msun}{{\mbox{M$_{\odot}$}}}
\newcommand{\kms}{\mbox{km\ s${^{-1}}$}}
\newcommand{\lya}{\mbox{${\rm Ly}\alpha$}}
\newcommand{\cmjj}{\mbox{${\rm cm^{-2}}$}}
\newcommand{\cmjjj}{\mbox{${\rm cm^{-3}}$}}

\shorttitle{QUEST Dwarfs: First release}
\shortauthors{Polzin et al.}

\begin{document}

\title[The Low-Mass Baryon Cycle in QUEST Dwarf Galaxies I: Sample definition and first results]{The Low-Mass Baryon Cycle in QUEST Dwarf Galaxies I: Sample definition and first results\vspace{-1.5cm}}

\author{Ava Polzin\,\orcidlink{0000-0002-5283-933X}$^{1,2\star}$}
\author{Hsiao-Wen Chen\,\orcidlink{0000-0001-8813-4182}$^{1,2}$}
\author{Zhijie Qu\,\orcidlink{0000-0002-2941-646X}$^{1,3}$}
\author{Erin Boettcher\,\orcidlink{0000-0003-3244-0409}$^{4,5,6}$}
\author{Allison Strom\,\orcidlink{0000-0001-6369-1636}$^{7,8}$}
\author{Tsang Keung Chan\,\orcidlink{0000-0003-2544-054X}$^{9,1}$\vspace{2mm}}
\affiliation{$^{1}$Department of Astronomy  \& Astrophysics, The University of Chicago, Chicago, IL 60637 USA}
\affiliation{$^{2}$Kavli Institute for Cosmological Physics, The University of Chicago, Chicago, IL 60637 USA}
\affiliation{$^{3}$Department of Astronomy, Tsinghua University, Beijing 100084, People’s Republic of China}
\affiliation{$^{4}$Department of Astronomy, University of Maryland, College Park, MD 20742, USA}
\affiliation{$^{5}$X-ray Astrophysics Laboratory, NASA/GSFC, Greenbelt, MD 20771, USA}
\affiliation{$^{6}$Center for Research and Exploration in Space Science and Technology, NASA/GSFC, Greenbelt, MD 20771, USA}
\affiliation{$^{7}$Department of Physics and Astronomy, Northwestern University, Evanston, IL 60208, USA}
\affiliation{$^{8}$Center for Interdisciplinary Exploration and Research in Astrophysics (CIERA), Northwestern University, Evanston, IL 60201, USA}
\affiliation{$^{9}$Department of Physics, The Chinese University of Hong Kong, Shatin, Hong Kong, People's Republic of China}

\thanks{$^\star$\href{mailto:apolzin@uchicago.edu}{apolzin@uchicago.edu}}

\begin{abstract}

We introduce the QUEST Dwarfs program, which connects low-mass galaxies' stellar populations, interstellar medium (ISM), and circumgalactic medium (CGM) in a large and coherently analyzed sample using a combination of optical spectroscopy, broadband imaging, and far ultraviolet (FUV) absorption spectra of bright background sources. 
We present initial results from the first-release sample, comprising 14 galaxies with stellar mass $\mstar\leq 10^9\,\msun$ at $z\approx0.001$--0.017, each with at least one CGM absorption probe at projected distances $d_{\rm proj}\lesssim 100$ kpc.
The sample broadly represents the field dwarf population and 
triples the number of available probes within 1/3 of the halo radius of dwarf galaxies outside of the Local Group. 
We find that the total silicon column density declines much more rapidly with projected distance than \ion{H}{1}, implying that chemically enriched cool gas is preferentially concentrated in the inner CGM, while the increasing ionization fraction of hydrogen with radius likely enhances this contrast. Accounting for unobserved silicon in higher ionization stages, we infer total metal masses of $\log M_Z/\msun\approx4.8$ and $6.5$ in the cool CGM within $0.3\,R_{\rm vir}$ for dwarfs with median $\log\,\mstar/\msun=7.6$ and 8.6, respectively. These reservoirs correspond to $\approx3$\% and $\approx16$\% of the total metals produced over the galaxies’ lifetimes.  
More massive galaxies also exhibit systematically stronger metal absorption, suggesting that projected distance governs the radial decline of metal absorption while stellar mass sets the normalization of the CGM metal profile. 
Individual ions reveal a multiphase structure, with low-ionization species concentrated toward the inner halo and higher-ionization species extending to larger distances.  Possible trends with recent star formation and gas-phase metallicity remain difficult to separate from their covariance with stellar mass. 
The full QUEST Dwarfs survey will provide the statistical power needed to isolate the dominant drivers of CGM enrichment in low-mass halos.

\end{abstract}


\section{Introduction}

The circumgalactic medium \citep[CGM; see, e.g.,][for recent reviews]{Tumlinson.Peeples.Werk.2017,Faucher-Giguere.Oh.2023,Chen.Zahedy.2026} is a critical component of galaxies' cosmic ecosystems. It serves as a reservoir of gas bound to a galaxy's halo, including both cool, neutral gas that, once accreted, will act as fuel for star formation \citep[e.g.,][]{Weng.etal.2023, Decataldo.etal.2024} and metal-enriched gas that was driven out of the galaxy by jets and wind \citep[e.g.,][]{Peeples.etal.2014, Muratov.etal.2017}. 

Because the multiphase (in temperature, chemistry, and kinematics) CGM acts as a reservoir for recently expelled gas, it encodes critical information about feedback processes occurring within the galaxy \citep[e.g.,][]{Shen.etal.2013, Suresh.etal.2015, Fielding.etal.2017, Medlock.etal.2025, Baumschlager.etal.2025}. It can be challenging to quantify the extent or the nature of the circumgalactic gas because it is generally diffuse, and emission studies are often limited to highly ionized phases in the inner CGM \citep[e.g.,][]{Augustin.etal.2019, Zhang.Zaritsky.2024, Peng.etal.2025} or hot halos that are X-ray bright \citep[e.g.,][]{WijersSchaye.2022, Silich.etal.2025}. Instead, the CGM can be traced in far ultraviolet (FUV) absorption \citep[e.g.,][]{Rauch.1998, Steidel.etal.2010, Chen.2017}, which provides access to a suite of metal resonance lines sensitive to different ionization conditions (included in this analysis: \ion{Si}{2}, \ion{Si}{3}, \ion{Si}{4}, \ion{C}{2}, \ion{C}{4}), as well as the hydrogen Lyman series. This method, however, relies on the chance alignment of UV-bright background sources with objects of interest. By necessity, galaxies' gaseous halos are then often characterized along one-dimensional ``pencil-beam'' sightlines.

\begin{deluxetable*}{l|ccccccccc}
    \tablecaption{Summary of first release QUEST Dwarfs galaxies and their properties. \label{tab:dat}}
    \tablewidth{\linewidth}
    \tablehead{\textbf{Galaxy}\tablenotemark{a}  &  \textbf{RA} & \textbf{Dec} & \textbf{\boldmath$D$ (Mpc)\unboldmath}\tablenotemark{b} &
    \textbf{\boldmath$cz \; (\mathrm{km\;s}^{-1})$\unboldmath}\tablenotemark{c} &
    \textbf{\boldmath$\log\,\mstar/\msun$\unboldmath}\tablenotemark{d} & \textbf{QSO} & \textbf{\boldmath$d_\mathrm{proj}$\unboldmath~(kpc)} & \textbf{Environment}\tablenotemark{e} \\
    (1)  & (2) & (3) & (4) & (5) & (6) & (7) & (8) & (9)}
    \startdata
    \textbf{PGC 4143} & 01:09:41.5 & $-$02:15:59.3 & $21\pm2$ & 1770$+109\atop-129$ & $8.9\pm0.02$ & LBQS 0107$-$0232 & $51\pm5$ & Isolated\\
     &  &  &  &  &  & LBQS 0107$-$0235 & $54\pm5$ & \\
     &  &  &  &  &  & QSO J0110$-$0218 & $56\pm5$ & \\
    \textbf{UGCA 285} & 12:33:08.5 & $-$00:31:55.0 & $9{+1\atop-2}$ & 754$+15\atop-12$ & $7.5{+0.2\atop-0.3}$ & LBQS 1230$-$0015 & $3.0{+0.3\atop-0.7}$ & Group\\
    \textbf{UGC 7370} & 12:19:40.3 & $+$02:04:46.5 & $35{+1\atop-2}$ & 1949$+92\atop-119$ & $9.4\pm0.2$ & QSO B1217$+$023 & $81{+2\atop-5}$ & Group\\
    \textbf{PGC 41458} & 12:31:33.4 & $+$12:03:49.6 & $9\pm2$& 596$+147\atop-150$ & $7.6\pm0.3$& LEDA 169472 & $13\pm3$ & Interacting\\
    \textit{PGC 41458b} & 12:31:34.0 & $+$12:04:03.4 &$10{+11\atop-6}$& 688$+314\atop-396$& $7.2{+0.9\atop-0.5}$&  &  & \\
    \textbf{LeG 26} & 10:51:21.1 & $+$12:50:56.0 & $8.9{+1.1\atop-0.9}$&481$+58\atop-52$& $7.7\pm0.2$& 2MASS J10512569$+$1247462 & $8.7{+1.1\atop-0.9}$ & Group\\
    \textbf{UGC 75} & 00:08:46.2 & $+$15:49:01.0 & $5\pm2$ & 877$+138\atop-97$ & $8.3\pm0.4$ & QSO B0003$+$1553 & $66\pm26$ & Interacting\\
    \textbf{UGC 9126} & 14:15:40.1 & $+$16:32:59.0 & $43\pm2$ & 2575$+97\atop-129$ & $9.2\pm0.2$ & SDSS J141542.90$+$163413.8 & $17.8\pm0.8$ & Isolated\\
    \textbf{UGC 7485} & 12:24:22.2 & $+$21:09:36.0 & $24.01\pm0.08$& $1183.0{+1.0 \atop -0.9}$ & $8.9\pm0.2$ & 4C21.35 & $105.9\pm0.3$ & Isolated\\
    \textbf{LEDA 1722581} & 08:35:37.1 & $+$25:00:15.0 & $80.55{+0.06\atop-0.11}$ & 5158$+4\atop-8$ & $8.8\pm0.2$ & FBQS J083535.8$+$245940 & $15.19{+0.01\atop-0.02}$ & Interacting\\
    \textit{LEDA 1722581-em} & 08:35:36.7 & $+$25:00:15.0 & $80.7{+0.4\atop-0.2}$ & 5166$+32\atop-14$ & $8.37\pm0.02$ &  &  & \\
    \textbf{PGC 1818175} & 12:17:08.2 & $+$27:47:42.9 & $16.9\pm0.1$\tablenotemark{f} & 971$+6\atop-15$ & $8.4\pm0.2$ & RX J1217.2+2749 & $12.99\pm0.08$ & Interacting\\
    \textbf{UGC 5427} & 10:04:41.0 & $+$29:21:51.5 & $10{+1\atop-2}$ & 447$+39\atop-18$& $8.0{+0.2\atop-0.3}$ & Ton 28 & $80{+8\atop-16}$  & Isolated\\
    \textbf{NGC 5486} & 14:07:25.0 & $+$55:06:10.5 & $25{+3\atop-4}$ & 1345$+35\atop-36$ & $9.2\pm0.2$ & QSO J1407$+$5507 & $12{+1\atop-2}$ & Group\\
    \textbf{dw1408$+$56} & 14:08:41.0 & $+$56:55:38.0 & $32.2{+0.9\atop-1.3}$ & 2163$+60\atop-97$ & $7.9\pm0.2$ & WISEA J140854.15$+$565743.3 & $25.8{+0.7\atop-1.0}$ & Isolated\\
    \textbf{UGC 4527} & 08:44:24.0 & $+$76:55:04.6 & $10.96\pm0.03$ & $725\pm4$ & $7.6\pm0.2$ & VII Zw 244& $7.23\pm0.02$ & Isolated
    \enddata
    \tablenotetext{a}{New and interacting satellites of QUEST Dwarfs, which are also analyzed in this work, have their names italicized. }
    \tablenotetext{b}{See \S\,\ref{sec:distance} for details regarding distance inference.}
    \tablenotetext{c}{See \S\,\ref{subsec:cz} for details regarding redshift measurements.}
    \tablenotetext{d}{See \S\,\ref{sec:mass} for details regarding measurements of \mstar.}
    \tablenotetext{e}{See \S\,\ref{subsec:environ} discussions on the environment of these dwarfs.}
    \tablenotetext{f}{The Cosmicflows-3 \citep{Kourkchi.etal.2020} model does not increase monotonically around $970$ km s$^{-1}$, so we adopt the mean of the potential model velocities.}
\end{deluxetable*}

Studying the CGM in absorption along pencil-beam sightlines is relatively advanced field for massive galaxies, with many systems examined and characterized. However, few such observational studies include dwarf galaxies \citep{Prochaska.etal.2011, Bordoloi.etal.2014, Liang.Chen.2014, Johnson.etal.2017, Johnson.etal.2026, Zheng.etal.2020, Zheng.etal.2024, Qu.Bregman.2022, Mishra.etal.2024, Fox:2026} with yet fewer focused on dwarfs outside of the Local Group \citep[e.g.,][]{McConnachie:2012AJ} at distances greater than 3 Mpc \citep{Prochaska.etal.2011, Bordoloi.etal.2014, Liang.Chen.2014, Johnson.etal.2017, Johnson.etal.2026, Zheng.etal.2024, Mishra.etal.2024} due to the intrinsic challenges related to their faintness and small sizes, which also affect the fruitful extent of the impact parameter. To get the strongest absorption, it is critical to analyze sightlines that are at small impact parameters \citep[ideally $\lesssim 0.5\, R_\mathrm{vir}$, e.g.,][]{Liang.Chen.2014}. Given typical dwarf galaxy halo masses \citep[e.g.,][]{Read.etal.2017, Nadler.etal.2020, Manwadkar.Kravtsov.2022}, sightlines must generally be within 50-100 kpc to stay within $R_\mathrm{vir}$, though \ion{H}{1} absorption is often extended and \ion{O}{6} absorption has been seen out to $2R_\mathrm{vir}$ \citep{Mishra.etal.2024}.

Additionally, there are indications that the CGM of dwarf galaxies is somewhat different than in more massive galaxies due to intrinsic differences in the baryon cycle for low-mass systems \citep{Li.etal.2021, Zhu.etal.2024}. It is more easily stripped, encodes different feedback processes, and has somewhat different chemistry. Understanding the CGM is critical to understanding feedback processes \citep{Baumschlager.etal.2025} and the baryon cycle \citep{Piacitelli.etal.2025} in this regime. The CGM is sensitive to feedback and formation/growth history, as, especially in dwarf galaxies, star formation itself can drive gas into the halo. In this vein, we combine absorption probes of the CGM content and constraints on interstellar medium (ISM) of the galaxies, stellar populations, and star formation histories to directly link the different components of the baryon cycle.

The initial Quasars to Understand Environments around, and STar formation in, Dwarfs -- or QUEST Dwarfs -- sample is made up of $>60$ dwarf galaxies with background FUV sightlines projected within $\sim R_\mathrm{vir}$. In this paper, we preview the science enabled by this large and diverse dataset, looking in detail at 14 systems, and highlight the ability to connect the chemical and physical properties of galaxies to the conditions of their CGM. All new data and data products presented here are available for easy download via the QUEST Dwarfs website\footnote{\href{https://questdwarfs.github.io}{https://questdwarfs.github.io}}, with archival data linked. 

In \S\  \ref{sec:data}, we describe the data used in this paper and how the larger QUEST Dwarfs sample was constructed. We detail the analyses of the first release sub-sample of galaxies presented here in \S\  \ref{sec:galanalysis}. In \S\,\ref{sec:cgmanalysis}, we expand on the analysis of the galaxies' larger environments, including measurements of their circumgalactic media. In \S\S\  \ref{sec:discussion} and \ref{sec:conclusions}, we discuss and then summarize our results.

\section{Sample Selection \& Data} \label{sec:data}

The QUEST Dwarfs are all selected from Cosmicflows-3 \citep{Tully.etal.2016}, the \citet[][]{Paudel.etal.2018} catalog of merging dwarf galaxies, and the 50 Mpc Galaxy Catalog \citep[50 MGC;][]{Ohlson.etal.2024} on the following criteria:
\begin{enumerate}
    \item low inferred stellar mass with $\mstar \leq 10^{9}\,\msun$ based on an initial mass-to-light ratio conversion given their broadband photometry and inferred distance in the literature (see \S\,\ref{sec:mass} for the mass inference);
    \item line-of-sight velocity distinguishable from the Milky Way -- $v \gtrsim 300$ km s$^{-1}$, which helps prevent confusion with absorption features due to high velocity Milky Way clouds; and
    \item available high-quality FUV absorption spectra of background source(s)\footnote{We will explore the subsample of systems that, like PGC 4143 introduced here, have multiple associated sightlines in a future work (Polzin et al. in preparation).} at small impact parameters ($d_\mathrm{proj}$), allowing for characterization of the CGM and galaxy's larger gaseous environment (see \S\,\ref{subsec:fuvspec} below).
\end{enumerate}

The combined QUEST Dwarfs sample contains $>60$ dwarf galaxies which span environments, star formation histories, and more than three decades in mass.

\begin{figure*}
    \centering
    \includegraphics[width = 0.75\linewidth]{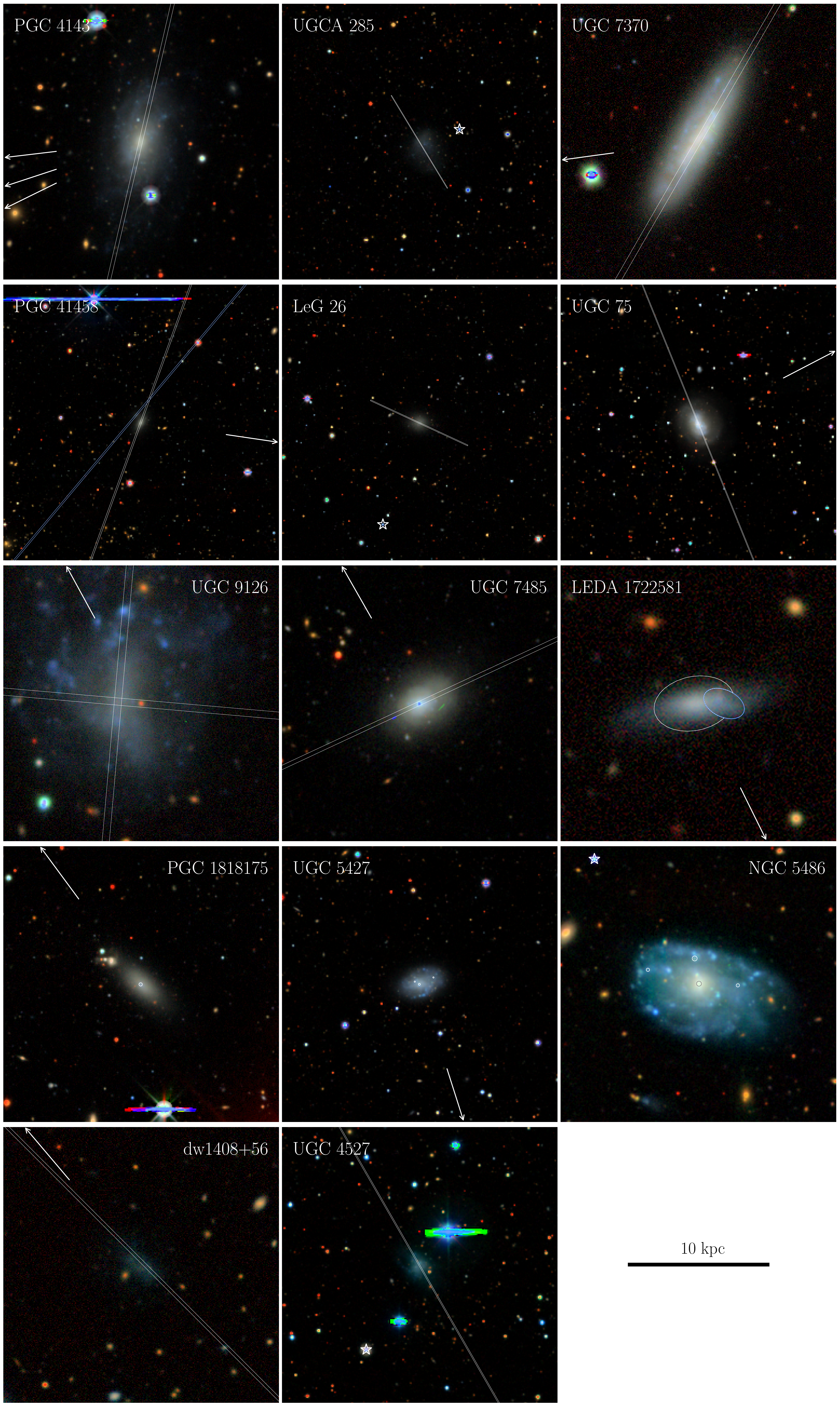}
    \caption{Cutouts of the QUEST Dwarfs first release sample from the Dark Energy Camera Legacy Survey \citep{Dey.etal.2019} in \textit{grz}. The scale bar applies to all postage stamps. QSOs in frame are marked with a star symbol; otherwise, arrows point toward the relevant sightline. The footprint of the slit, fiber, or IFU region used for optical spectroscopy is overplotted on each galaxy. The ancillary targets, PGC 41458b and LEDA 1722581-em, are marked by a blue slit and an extraction region, respectively. The central pointing of NGC 5486, which is largely excluded from analyses because of strong P Cygni-like features, is shown in gray.
    \label{fig:postagestamps}}
\end{figure*}

The sub-sample analyzed here consists of 14 galaxies with available high-quality multi-wavelength imaging and spectroscopic data, either retrieved from public archives or obtained ourselves. It showcases the range of science possible with these combined datasets. A gallery of color composite images of these galaxies are presented in Figure \ref{fig:postagestamps}, while the galaxy name and coordinates are presented in columns (1) through (3) of Table \ref{tab:dat} which also lists the corresponding QSO sightline and projected distance ($d_\mathrm{proj}$) in columns (7) and (8).

Combining optical spectroscopy with broad-band photometry provides constraints on the galaxies’ stellar populations, star formation histories, and ISM metallicities. These complementary data enable a unified study of the low-mass baryon cycle by linking the stellar populations, ISM, and CGM within the same galaxy sample. In this section, we briefly describe the available observations. A summary of the optical imaging and spectroscopic data is presented in Table \ref{tab:highres}. The remaining QUEST Dwarfs will be presented in future data releases, with accompanying observations and derived data products made publicly available as each sample is published.
   
\subsection{Optical Galaxy Spectra}

As summarized in Table \ref{tab:highres}, optical spectra for eight dwarf galaxies in the first release sample were obtained ourselves, while the remaining six galaxies have existing data in either the Keck archive or from the Sloan Digital Sky Survey \citep[e.g.,][]{York:2000}\footnote{The final flux-calibrated, median-combined 1-D spectra from this paper are available via the QUEST Dwarfs website and GitHub.}.  

\begin{deluxetable}{l|ccc}
\tablecaption{Summary of spectral and imaging observations of QUEST Dwarfs. \label{tab:highres}}
    \tablewidth{\linewidth}
    \tablehead{\textbf{Galaxy}  &  \textbf{\textit{HST}/WFPC2} & \textbf{\textit{HST}/ACS} & \textbf{Optical spectrum}}
    \startdata
    \textbf{PGC 4143} & - & - & Magellan/IMACS \\
    \textbf{UGCA 285} & - & F814W & Keck/LRIS \\
    \textbf{UGC 7370} & - & - & Magellan/IMACS\\
    \textbf{PGC 41458} & - & - & Magellan/IMACS\\
    \textbf{LeG 26} & - & - & Keck/LRIS\\
    \textbf{UGC 75} & F300W, F814W & - & Magellan/IMACS\\
    \textbf{UGC 9126} & - & - & Magellan/IMACS\\
    \textbf{UGC 7485} & - & - & Magellan/IMACS\\
    \textbf{LEDA 1722581} & - & F814W & SDSS MaNGA\\
    \textbf{PGC 1818175} & - & - & SDSS\\
    \textbf{UGC 5427} & - & F606W, F814W & SDSS\\
    \textbf{NGC 5486} & - & F555W, F814W & SDSS\\
    \textbf{dw1408$+$56} & - & - & MMT/Binospec\\
    \textbf{UGC 4527} & - & - & MMT/Binospec
    \enddata
\end{deluxetable}

\subsubsection{Magellan/IMACS}
We observed UGC 75 on the Inamori-Magellan Areal Camera and Spectrograph \citep[IMACS;][]{Dressler.etal.2011} on Magellan I on 2022 November 15, taking three exposures of UGC 75 of 600 s, 1800 s, and 1200 s respectively, totaling 3600 s. We also observed PGC 4143 with IMACS on 2022 November 16 for a total of 2400 s across two 1200 s exposures. UGC 7370 was observed with IMACS for 2400 s (2$\times 1200$ s) on 2023 March 22. UGC 9126 was observed on both 28 April and 2 May 2025, integrating for 3600s ($3 \times 1200$s) each day -- the slit was oriented along the galaxy's major axis on 28 April and along the galaxy's minor axis on 2 May. PGC 41458 and its smaller companion, which we will refer to here as PGC 41458b, were each observed on 2 May 2025 for a total of 3600s ($3 \times 1200$ s). These spectra were all taken with IMACS's f/4 channel using the 2\farcs5-wide long slit to maximize signal from our low surface brightness targets and the 300 l mm$^{-1}$ grating blazed at 6650 \AA. 
Nominally, this grating covers from $\sim3700-9700$\AA\ 
with a spectral FWHM of $\sim$17 \AA~(or $R \sim 390$). 

UGC 7485 was observed on 2024 11 February using the f/2 channel on IMACS with the 2\farcs5 long slit and the 400 l mm$^{-1}$ grism blazed at 4730 \AA. This instrument setup has a similar wavelength range to the other IMACS observations used here, covering 3900-8500 \AA~ at $R\sim420$ (FWHM $\sim 11$ \AA), though we only use data from the redder chip in this analysis with wavelengths longer than 4500 \AA.

In each case, with the exception of the second UGC 9126 pointing on 2 May 2025, the slit position angle was aligned with the major axis of the galaxy to maximize coverage of any spatial variation in the resulting spectra.

We reduce these data using a proprietary \texttt{IDL}-based pipeline built for Magellan analysis. Using the sky-subtracted spectra, we create a combined 2D spectral image for each target on a by-pointing basis, taking the inverse variance-weighted average for two exposures or median for three or more exposures, and mask cosmic rays and artifacts from the sky subtraction. We define our boxcar spectral extraction region to include the full spatial extent of the galaxies, so that the resultant one-dimensional spectrum is summed across the extended stellar body of the target with wavelength corrected to vacuum.  Therefore, the recovered properties based on these longslit spectra in \S\,\ref{sec:galanalysis} are integrated over the galaxy.

\subsubsection{MMT/Binospec}

dw1408$+$56 and UGC 4527 were both observed with MMT's Binospec optical spectrograph \citep{Fabricant.etal.2019} using the 1\farcs5 single slit mode with the 270 l mm$^{-1}$ grating blazed at 6500 \AA, which covers roughly 3850--9150 \AA. The slit was aligned with the major axis of each target galaxy to maximize spatial coverage. No additional binning was applied and the science frames have a spectral dispersion of 1.3 \AA~pix$^{-1}$ and a spatial resolution of 0.24 arcsec pix$^{-1}$.

Binospec, like other MMT instruments, uses queue scheduling. UGC 4527 was observed on 2025 March 31 for a total of 3600s across three 1200s exposures. dw1408$+$56 was observed on 2025 April 27 for a total of 4800s across four 1200s exposures. 

We independently extract the trace of UGC 4527 from the co-added and sky-subtracted, flux-calibrated two-dimensional spectra produced by MMT's automated Binospec pipeline \citep{Chilingarian.etal.2015}. Due to anomalous issues with the automatic wavelength calibration on the night of 27 April 2025, we reran the MMT/Binospec pipeline\footnote{\href{https://bitbucket.org/chil_sai/binospec/src/master/}{https://bitbucket.org/chil\_sai/binospec/src/master/}} for dw1408$+$56. We largely use the pipeline's default parameters, but we turn off the sky line illumination correction and, given the low surface brightness of dw1408$+$56, the ``bright'' source flag.  

\subsubsection{Keck/LRIS}

UGCA 285 was previously observed for a total of 600\,s with the Keck I's Low Resolution Imaging Spectrograph \citep[LRIS;][]{Oke.etal.1995} on 2016 January 9 under program U050LA (PI: Prochaska). Data were taken with the 1\arcsec~long slit using the 400 l mm$^{-1}$ grism blazed at 3400\AA~(blue) and the 600 l mm$^{-1}$ grating blazed at 7500\AA~(red) with the D560 dichroic, covering 1760--8400\AA. The slit was angled along the galaxy's major axis with PA = 31.2 deg. Collected data underwent $2\times2$ binning, resulting in a spatial resolution of $0.27$ arcsec pix$^{-1}$ and a spectral dispersion of 2.18 \AA~pix$^{-1}$ (blue) and 1.6 \AA~pix$^{-1}$ (red).

LeG 26 was observed with Keck/LRIS on 2003 April 26 for program U21L (PI: Bolte) using the 1\farcs5 long slit. Data are not binned spatially ($0.135$ arcsec pix$^{-1}$) or spectrally and cover from approximately 3040--8630\AA, with the 600 l mm$^{-1}$ grism blazed at 4000\AA~(blue, 0.63 \AA~pix$^{-1}$), the 600 l mm$^{-1}$ grating blazed at 7500\AA~(red, 0.80 \AA~pix$^{-1}$), and the D560 dichroic. There are three separate exposures of 900\,s each.

We reduced these data, obtained via the Keck Observatory Archive, with \texttt{PypeIt} \citep{Prochaska.etal.2020}, using the default parameters for LRIS reduction with the following exceptions: we turn off automatic trace extraction due to the low surface brightness of these sources and define both a broad FWHM and low SNR for source detection to avoid oversubtracting signal during the sky modeling step. On the blue side, there are no lamp-on flats from the relevant night for UGCA 285, so pixel-level flat-fielding is not completed.

\begin{deluxetable*}{l|cccccc}
    \tablecaption{Details of QSO sightlines and their spectral coverage \label{tab:qsos}}
    \tablewidth{\linewidth}
    \tablehead{\textbf{QSO}  &  \boldmath\textbf{$z_\mathrm{QSO}$}\unboldmath &   \textbf{COS Grating} & \boldmath\textbf{$\langle\,S/N\,\rangle_{\rm COS}$}\unboldmath &    \textbf{COS PID} & \boldmath\textbf{$\langle\,S/N\,\rangle_{\rm FUSE}$}\unboldmath & \textbf{FUSE PID}}
    \startdata

    \textbf{LBQS 0107-0232} & 0.727 & G160M & 10.3 & 11585 
    & - & - \\
\textbf{LBQS 0107-0235} & 0.957 & G130M, G160M & 12.8 & 11585 
& - & - \\
\textbf{QSO J0110-0218} & 0.956 & G130M, G160M & 11.5 & 11585 
& - & - \\
\textbf{LBQS 1230-0015} & 0.471 & G130M, G160M & 10.5 & 11598, 12486 
& - & - \\
\textbf{QSO B1217+023} & 0.240 & G130M, G160M & 13.5 & 13852 
& - & - \\
\textbf{LEDA 169472} & 0.116 & G130M & 12.7 & 14071 
& & - \\
\textbf{2MASS J10512569+1247462} & 1.281 & G130M, G160M & 6.0 & 12603, 14777 
& - & - \\
\textbf{QSO B0003+1553} & 0.450 & G130M, G160M & 21.9 & 12038 
& - & - \\
\textbf{SDSS J141542.90+163413.8} & 0.743 & G130M & 15.8 & 12486 
& - & - \\
\textbf{4C21.35} & 0.432 & G130M, G160M & 20.6 & 13008 
& - & - \\
\textbf{FBQS J083535.8+245940} & 0.331 & G130M, G160M & 12.3 & 12025 
& 3.1 & P207 \\
\textbf{RX J1217.2+2749} & 0.395 & G130M & 8.0 & 14772 
& - & - \\
\textbf{Ton 28} & 0.330 & G130M, G160M & 19.5 & 12038 
& 2.5 & P207\tablenotemark{a} \\
\textbf{QSO J1407+5507} & 1.027 & G130M & 12.0 & 12486 
& - & - \\
\textbf{WISEA J140854.15+565743.3} & 0.336 & G130M & 11.7 & 13314 
& - & - \\
\textbf{VII Zw 244} & 0.131 & G130M, G160M & 21.6 & 11520
& 8.7 & G020 

    \enddata
\tablenotetext{a}{These data are not used in our analysis due to low SNR.}
\end{deluxetable*}

We extract the trace using an unweighted boxcar, which we flux calibrate manually rather than using the \texttt{PypeIt} ``fluxing'' step. We mask bad pixels in the two-dimensional spectra and median combine the three LeG 26 exposures before recovering two one-dimensional spectra per galaxy, one from the blue side of the spectrograph and one from the red. The individual blue and red 1D spectra are then concatenated to form a final science spectrum joined according to the considerations described in \citet[][]{Perley.2019}. We note that the LeG 26 blue-side flux calibration is somewhat uncertain and necessitated rescaling to match the continuum level ascertained from the red detector; while no rescaling is necessary between red and blue detectors for UGCA 285, a slight bias correction is applied to its associated blue detector data.

\subsubsection{SDSS}

PGC 1818175, UGC 5427, and NGC 5486 have existing optical spectra taken as part of the Sloan Digital Sky Survey \citep[SDSS;][]{Blanton.etal.2017} at Apache Point Observatory. The SDSS spectrograph (and BOSS spectrograph) fibers are 3\arcsec (2\arcsec) in diameter and often target brighter, star-forming or \textsc{Hii} regions in the galaxies, making it hard to assess the overall properties of the galaxies. As a means of mitigating this lack of spatial averaging, we analyze all available SDSS spectra from multiple positions, taking the sum of line strengths and the 
luminosity-weighted 
average of fit properties derived from the different pointings where possible to ascertain the \textit{galaxy's} more global properties.

PGC 1818175 was observed once (spectroscopic object ID 2511998754664507392), targeting the central region of the galaxy. The fiber covers the bluest, most apparently star-forming portion of the galaxy.

UGC 5427 has three SDSS spectra available from different pointings. Spectroscopic object ID 2195517795251283968 encompasses a star-forming region near the center of the galaxy, ID 7283574696637126656 is relatively central, but not focused on a specific bright clump, and ID 12779137532475758592 is pointed at the galaxy's outskirts.

There are four SDSS spectroscopic object IDs associated with NGC 5486. The fiber for spectroscopic ID 1490826493520013312 targeted the center of the galaxy, but the resulting spectrum shows broad P Cygni-like line profiles that are not easily fit by a generic model. In this analysis, we use IDs 9489301150144681984, 9488057877522569216, and 1491909507004000256, which each target one of the galaxy's brighter star-forming regions. 

LEDA\,1722581\footnote{This galaxy has a massive neighbor, NGC\,2611, at $d_{\rm proj}=60$ kpc from the QSO sightline based on an updated distance (see \S\,\ref{sec:distance}).  NGC\,2611 is, therefore, four times farther away from the QSO sightline than LEDA\,1722581. The absorption features identified in the spectrum of FBQS J083535.8$+$245940 (\S\,\ref{sec:cgmanalysis} and Table \ref{tab:ion}) were previously attributed to NGC\,2611 \citep{Stocke.etal.2013}. However, following the association criterion described in \S\,\ref{subsec:environ}, we consider LEDA\,1722581 the more likely host because of its substantially closer proximity to the QSO sightline.} at $d_{\rm proj}=15$ kpc to the sightline of FBQS J083535.8$+$245940 
was observed as part of the Mapping Nearby Galaxies at Apache Point Observatory \citep[MaNGA;][]{Bundy.etal.2015} integral field survey (MaNGA ID 1-385149), which bundles BOSS fibers to simultaneously gather spatial and spectral information about targeted galaxies. We use data cubes processed with the MaNGA Data Analysis Pipeline \citep{Westfall.etal.2019, Belfiore.etal.2019} via \texttt{Marvin} \citep{Cherinka.etal.2019} to access the spaxel-level H$\alpha$ line intensity map. We then segment the object into two different regions, one that is cospatial with the stellar continuum, and one that is slightly offset and appears primarily in H$\alpha$ and \textit{GALEX} FUV emission, which we refer to as LEDA 1722581-em here. We analyze the co-added spectra from each of these regions separately.

\subsection{FUV QSO Spectra}
\label{subsec:fuvspec}

Each of the dwarf galaxies in the QUEST Dwarf sample has at least one background quasar at projected separation $d_\mathrm{proj} \lesssim 100$ kpc at their inferred distance, for which there exist archival FUV absorption spectra obtained using the Cosmic Origins Spectrograph \citep[COS;][]{COS} on board the \textit{Hubble Space Telescope} (\textit{HST}). Three of the QSOs also have archival FUV spectra obtained using the \textit{Far Ultraviolet Spectroscopic Explorer} \citep[\textit{FUSE};][]{FUSE}.  A summary of these FUV spectra is presented in Table \ref{tab:qsos}. \textit{FUSE} spectra cover a wavelength range from 905 to 1187 \AA\ with a full-width-at-half-maximum spectral resolution of ${\rm FWHM}\approx 15\,\kms$ in line-of-sight velocity.  \textit{HST}/COS spectra typically cover a wavelength range from 1140 to 1780 \AA\ with a velocity resolution of ${\rm FWHM}\approx 18\,\kms$.  Together, these FUV spectra provide spectral coverage for a suite of \ion{H}{1} and ionic transitions that are necessary for constraining the chemical and physical conditions of the diffuse CGM \citep[e.g.,][]{Chen.Zahedy.2026}. 
The extended wavelength coverage from \textit{FUSE} is particularly useful for constraining the neutral hydrogen column density, $N({\rm H\textsc{I}})$, based on the strengths of higher-order Lyman series absorption lines.

Both FUSE and \textit{HST}/COS spectra were processed and co-added using custom software we developed to account for wavelength-calibration errors and to optimize the signal-to-noise ($S/N$) in the final combined spectra. Details regarding data processing of the FUSE and \textit{HST}/COS spectra are described in \cite{Qu:2026} and \cite{Chen:2020}, respectively.  The median $S/N$ of the coadded FUSE spectra along three QSO sightlines ranges from $\langle\,S/N\,\rangle_{\rm med}\approx 2.5$ to $\langle\,S/N\,\rangle_{\rm med}\approx 8.7$ per resolution element, while the median $S/N$ of individual coadded \textit{HST}/COS spectra ranges from $\langle\,S/N\,\rangle_{\rm med}\approx 6$ to $\langle\,S/N\,\rangle_{\rm med}\approx 22$ per resolution element \citep[see][and Table \ref{tab:qsos} for details]{Qu:2026}.

\subsection{Galaxy Images}

The QUEST Dwarfs are covered by multiwavelength archival data from a variety of sources, allowing us to directly characterize the morphology, stellar populations, and recent star formation history of the galaxies in this sample.

\subsubsection{Optical Survey Imaging}

All of the QUEST Dwarfs were observed by either the Dark Energy Camera Legacy Survey \citep[DECaLS;][]{Dey.etal.2019} or the complementary ancillary imaging in the DECam Local Volume Exploration Survey \citep[DELVE;][]{Drlica-Wagner.etal.2021}. The 14 primary galaxies analyzed here are all included in DECaLS (see Figure \ref{fig:postagestamps}), and we use DR9 $grz$ ``bricks'' with approximately native resolution (0.262\arcsec~pix$^{-1}$) to derive the integrated photometric properties of each dwarf galaxy.

\subsubsection{High-resolution Imaging}

About half of the QUEST Dwarfs have archival high-resolution imaging, which enables detailed study of the galaxies' stellar populations. Of this first release sub-sample, five galaxies have space-based imaging taken with the \textit{Hubble Space Telescope} (PIDs 9124, 12546, 13024, and 17070; see Table \ref{tab:highres}). 

We download CTE-corrected individual exposure image files from the Barbara A. Mikulski Archive for Space Telescopes, which we ``tweak'', co-add, and ``drizzle'' via \texttt{Drizzlepac} \citep{Fruchter.etal.2010,Hoffmann.etal.2021}. This allows us to derive a consistently drizzled PSF for use with the resulting co-added and mosaiced images 
starting from the same data and processing.

\subsubsection{Ultraviolet Imaging}

Most of the QUEST Dwarfs are covered by \textit{GALEX} \citep{Martin.etal.2005} ultraviolet imaging. We use these data to derive a self-consistent star formation rate (SFR) on $\sim100$ Myr timescales \citep[e.g.,][]{Kennicutt.1998, Calzetti.2013, Flores-Velazquez.etal.2021}. With our optical spectra enabling direct measurement of the SFR on $\lesssim 10$ Myr timescales via the H$\alpha$ line, these ultraviolet data give us the opportunity to probe the recent star formation history of dwarf galaxies. The expectation is that star formation in this mass regime is generally feedback-regulated and somewhat bursty, which makes the combined dataset particularly valuable for constraining star formation properties. Simultaneously, each star formation indicator comes with caveats for use in the extremely low-mass, low-metallicity regime, which includes many of the QUEST Dwarfs: some particularly low-mass dwarf galaxies may not exhibit H$\alpha$ emission due to the intrinsic stochasticity in sampling the high-mass end of the stellar initial mass function in particularly low-mass galaxies \citep[e.g.,][]{Applebaum.etal.2020}, and low-metallicity galaxies tend to be bluer and more UV-bright regardless of their SFR \citep[e.g.,][]{Kennicutt.Evans.2012, Sanders.etal.2013}. It is therefore important to examine the recent star formation history of dwarf galaxies in greater detail and using multiple tracers.

\section{Properties of QUEST Dwarf Galaxies} \label{sec:galanalysis}

To better understand the correspondence between galaxy and halo absorber properties, we analyze available images and spectra to infer key physical and chemical quantities for these dwarf galaxies, including broad-band photometry, radial velocity, distance, stellar metallicity, star formation history, stellar mass, ISM metallicity, and environment.
Here we describe the steps for inferring these quantities.

\subsection{Integrated Photometric Properties of Galaxies}
\label{sec:photometry}

All of the galaxies in this sub-sample are covered by DECaLS \citep{Dey.etal.2019} DR9 imaging, and we fit for each galaxy's structural and photometric properties in \textit{grz} filters.  We use cube-by-cube \texttt{PSFEx} \citep{Bertin.2011, Bertin.2013} PSF models for Legacy Survey photometry.

To mask non-target galaxy features in the images, we use segmentation maps from \texttt{SEP} \citep{Bertin.Arnouts.1996, Barbary2016}, and run \texttt{pysersic} \citep{Pasha.Miller.2023} to fit simple S\'{e}rsic models jointly to the galaxies' $grz$-band light profiles. In the couple of cases where the galaxies' morphologies necessitate more complex modeling than a single S\'{e}rsic profile, we use a ``double'' S\'{e}rsic model, where only the profiles' centers and position angles are shared. These structural parameters are then used in aperture photometry of the deepest available \textit{GALEX} FUV imaging, defining the ellipse from which we extract each galaxies' FUV flux based on $\gtrsim2\times$ the effective radius recovered from the \texttt{pysersic} for to DECaLS imaging. Our targets are covered by both the main \textit{GALEX} surveys and guest investigator programs, including PIDs 25, 40, and 104.
All integrated photometry is reported in columns (1) through (4) of Table \ref{tab:phot}, along with the best-fitting morphological parameters of a single S\'{e}rsic model in columns (5) through (8) for each galaxy.

For the five galaxies also covered by \textit{HST}, we model the galaxies'  light profiles more carefully with \texttt{GALFIT} \citep{Peng.etal.2010}. We use \texttt{spike} \citep{Polzin.2025} with \texttt{TinyTim} \citep{Krist.etal.2011} to generate resampled instrument- and filter-dependent \textit{HST} point spread functions at the locations of UGCA 285, UGC 75, LEDA 1722581, UGC 5427, and NGC 5486. The drizzled \textit{HST} PSFs are then fed to \texttt{GALFIT}. Because these galaxies stellar populations' are resolved, structural parameters are first fit for \textit{HST} imaging smoothed with a Gaussian kernel and integrated magnitudes are then force fit for the original (unsmoothed) imaging.

\subsection{Radial Velocities of Galaxies} \label{subsec:cz}

To ensure the accuracy of the wavelength calibration we re-zeropoint the wavelength axis based on the location of the atmospheric A-band absorption feature in our newly extracted spectra, which results in a 2\AA~correction applied globally to the extracted LeG 26 spectrum. For our broad 2.5\arcsec~longslit data, we also correct for spatial variation within the slit, offsetting the wavelength array so that the measured $cz$ accounts for the potentially irregular line profiles associated with structure in the observed galaxies. All such corrections are at the $\lesssim 1$\AA~level, while the FWHM of the corrected data varies between $\sim$11\textendash17\AA.

We then measure the radial velocity of each galaxy in this sample using available optical spectra and a simple reduced $\chi^2$ minimization scheme. Template spectra are \texttt{FSPS} \citep{Conroy.Gunn.White.2009, Conroy.Gunn.2010, Johnson.etal.2024} simple stellar populations (SSPs) spanning both metallicity ($-2.5 \leq \mathrm{[Fe/H]} \leq 0$) and age ($0.5 - 10$ Gyr) and smoothed to approximate each instrument's spectral resolution. We report the median best fit velocity across templates with a heliocentric correction applied and confidence intervals defined by the 16th and 84th percentiles of the distribution of velocities returned across templates.

Where there are independent spectra at multiple pointings associated with the same galaxy, we take the average best fit velocity across pointings as the ``integrated'' $cz$. For UGC 9126, the spectrum taken along the minor axis is much noisier than the spectrum taken along the major axis; we adopt the major axis velocity.

In the few cases where no good fit to the (largely Balmer) absorption lines is possible and no alternative pointing exists, we use full spectral fitting (see \S\  \ref{subsubsec:sed}) to measure the galaxy's $cz$.  The best-fit $cz$ of each galaxy is presented in column (5) of Table \ref{tab:dat}.

\subsection{Inferring Distances}
\label{sec:distance}

We infer the distance to every galaxy presented here from either its radial velocity or, where possible, a redshift-independent distance calibrator. Most galaxies in the first release QUEST Dwarfs sample do not have redshift-independent distances from either the Tip of the Red Giant Branch \citep[TRGB; e.g.,][]{Lee.Freedman.Madore.1993} or surface brightness fluctuations \citep[SBFs;][]{Tonry.Schneider.1988}. For the subset with \textit{HST} imaging, we measure the SBF distance. The best estimated distances are presented in column (4) of Table \ref{tab:dat}.

\subsubsection{Radial Velocity Distances}

We use the Cosmicflows-3 Distance-Velocity Calculator \citep{Kourkchi.etal.2020} to compute local velocity field- and cosmology-corrected distances from galaxy radial velocities. We adjust the output distances to use $H_0 = 70\;\mathrm{km~s^{-1}~Mpc^{-1}}$ in lieu of the higher $H_0 = 75\;\mathrm{km~s^{-1}~Mpc^{-1}}$ used by default in the Cosmicflows-3 model.

Where possible, however, we adopt redshift-independent TRGB or SBF distances, whether from the literature or newly measured here, for the galaxies in this sample. For galaxies within $cz \lesssim 1500\,\mathrm{km~s^{-1}}$, the galaxy's peculiar velocity and any potential group velocity represent a significant source of uncertainty on the inferred radial velocity distance, as these velocity components may dominate over the Hubble flow. 

\subsubsection{Surface Brightness Fluctuation Distances}\label{subsubsec:sbf}

Four of the dwarf galaxies presented here have archival \textit{HST}/ACS F814W imaging, which allows us to apply existing surface brightness fluctuation distance calibrations \citep{Blakeslee.etal.2010, Carlsten.etal.2019b} that have been shown to recover accurate distances for younger stellar populations to put a redshift-independent constraint on the galaxies' distances. 

We find and apply a conversion between $(g-z)^\mathrm{DECam}$ and F475W-F814W using the colors of simple stellar populations in MIST as was done in, e.g., \citet{Cohen.etal.2018} and \citet{Carlsten.etal.2019b}. More details of this conversion are in Appendix \ref{app:SBF}.

For UGC 75, which has archival multi-band \textit{HST}/WFPC2 imaging, we use the F814W SBF relation from \citet[][]{Ajhar.etal.1997}. Though \citet[][]{Ajhar.etal.1997} calibrated the F814W SBF signal for PC imaging, this relation has been shown to work for WF imaging, as well \citep{Morris.Shanks.1998}. As above, we find a conversion between $(g-z)^\mathrm{DECam}$ and the standard $V-I$ (F555W$-$F814W) color on which that relation relies.

From the TRGB, UGC 5427 was previously measured to be at $7.7\pm0.77$ Mpc \citep{Tully.etal.2013}. We verify this distance, deriving a SBF distance of $10{+1\atop-2}$ Mpc from the F814W \textit{HST}/ACS imaging. For consistency with the treatment of the rest of our sample, we adopt the SBF distance in our analysis. 

Results are shown in Table \ref{tab:hstphot}. For the galaxies imaged with \textit{HST}/ACS, we adopt the \citet[][]{Carlsten.etal.2019b} SBF relation given the blue color of the target galaxies, but note that distance results from \citet[][]{Blakeslee.etal.2010} are consistent with those reported here. LEDA 1722581's variance power spectrum is background-dominated at all scales \citep[using the updated SBF equation from][]{Foster.etal.2024}; as there is no way to recover the SBF signal, it is not included in our reported results. Fortunately, LEDA 1722581 is in the regime where the galaxy should be securely in the Hubble flow.

\begin{deluxetable}{l|ccc}
    \tablecaption{SBF distance measurements \label{tab:hstphot}}
    \tablewidth{\linewidth}
    \tablehead{\textbf{Galaxy}  &  \boldmath$\bar{M}$\unboldmath & \boldmath$m_\textbf{SBF}$\unboldmath & \boldmath$D_\textbf{SBF}$\unboldmath~\textbf{(Mpc)}}
    \startdata
    \textbf{UGCA 285} & $-2.2\pm0.3$ & $27.5{+0.1\atop-0.3}$ & $9{+1\atop-2}$ \\
    \textbf{UGC 75} & $-3\pm1$ & $25.22{+0.06\atop-0.18}$ & $5\pm2$\\
    \textbf{UGC 5427} & $-2.3\pm0.2$ & $27.7{+0.2\atop-0.3}$ & $10{+1\atop-2}$\\
    \textbf{NGC 5486} & $-2.5\pm0.2$ & $29.48{+0.05\atop-0.20}$ & $25{+3\atop-4}$ 
    \enddata
\end{deluxetable}

\subsection{Spectral Fitting} \label{subsubsec:sed}

\begin{deluxetable*}{l|cccccccc}
    \tablecaption{Derived spectral and photometric properties. \label{tab:derived}}
    \tablewidth{\linewidth}
    \tablehead{\textbf{Galaxy}  & \textbf{\boldmath$\log M_\mathrm{h}/\msun$\unboldmath} & \textbf{\boldmath$R_\mathrm{vir}\;(\mathrm{kpc})$\unboldmath} & \textbf{\boldmath $\log Z/Z_\odot$\unboldmath}\tablenotemark{a} & \textbf{\boldmath $12 + \log \mathrm{O/H}$\unboldmath} & \textbf{\boldmath $\log$ sSFR$_\mathrm{H\alpha}$/yr$^{-1}$ \unboldmath} & \textbf{\boldmath $\log$ sSFR$_\mathrm{UV}$/yr$^{-1}$ \unboldmath} & 
    \textbf{\boldmath$\log \mathrm{H\alpha/UV}$\unboldmath} & \textbf{SFH}\tablenotemark{b} \\
    (1)  & (2) & (3) & (4) & (5) & (6) & (7) & (8) & (9)}
    \startdata
    \label{tab:derived}
    \textbf{PGC 4143} & 10.90 & 88.87 & $-0.953{+0.010 \atop -0.003}$ & $8.40\pm0.18$ & $-10.595\pm0.005$ & $-10.6\pm0.2$ & $0.0\pm0.2$ & B \\
    \textbf{UGCA 285} & 10.16 & 50.12 & $-0.07{+0.05\atop-0.08}$ & $8.62\pm0.19$ & $-10.07\pm0.05$ & $-10.6{+0.2\atop-0.4}$ & $0.5{+0.2\atop-0.4}$ & S \\
    \textbf{UGC 7370} & 11.13 & 105.59 & $-1.40{+0.05\atop-0.06}$ & $8.41\pm0.18$ & $-10.060\pm0.005$ & $-10.5\pm0.2$ & $0.5\pm0.2$ & B \\
    \textbf{PGC 41458} & 10.21 & 52.03 & $-1.63{+0.11\atop-0.08}$ & $8.56\pm0.40$ & $-12.464\pm0.003$ & $-11.7{+0.2\atop-0.4}$ & $-0.7\pm0.4$ & Q \\
    \textit{PGC 41458b} & 10.21 & 45.02 & $-0.009{+0.006\atop-0.014}$ & $8.83\pm0.14$\tablenotemark{c} & $-12.3{+0.7\atop-2.2}$ & $<-11.94$ & $>-0.33$ & Q \\
    \textbf{LeG 26} & 10.25 & 54.04 & $-0.05{+0.03\atop-0.06}$ & $8.31\pm0.19$ & $-11.15{+0.06\atop-0.097}$ & $-11.6\pm0.2$ & $0.5{+0.2\atop-0.5}$ & Q\\
    \textbf{UGC 75} & 10.58 & 69.28 & $-1.45{+0.05\atop-0.06}$ & $8.37\pm0.18$ & $-10.195\pm0.002$ & ... & ... & S \\
    \textbf{UGC 9126} & 11.04 & 98.79 & $-0.99\pm0.05$ & $8.14\pm0.18$ & $-9.58\pm0.01$ & $-9.9\pm0.2$ & $0.3\pm0.2$ & E\\
    \textbf{UGC 7485} & 10.90 & 88.87 & $-0.72{+0.06\atop-0.02}$ & $8.29\pm0.18$  & $-9.460{+0.008\atop0.009}$ & $-10.5\pm0.2$ & $1.1\pm0.2$ & S \\
    \textbf{LEDA 1722581} & 10.85 & 85.55 & $-1.19\pm0.04$ & $8.36\pm0.18$ & $-10.300\pm0.004$ & $-10.8\pm0.2$ & $0.5\pm0.2$ & S \\
    \textit{LEDA 1722581-em} & 10.62 & 71.41 & $-1.26{+0.04\atop-0.06}$ & $8.31\pm0.18$ & $-9.722\pm0.004$ & $-10.01\pm0.08$ & $0.29\pm0.08$ & S \\
    \textbf{PGC 1818175} & 10.64 & 72.39 & $-0.93{+0.05\atop-0.06}$ & $8.86\pm0.35$ & $-11.7\pm0.1$ & $-11.8{+0.3\atop-0.2}$ & $0.1{+0.3\atop-0.2}$ & Q \\
    \textbf{UGC 5427} & 10.41 & 60.93 & $-1.30\pm0.02$ & $8.28\pm0.18$  & $-9.68\pm0.01$ & $-9.8{+0.2\atop-0.3}$ & $0.1{+0.2\atop-0.3}$ & E \\
    \textbf{NGC 5486} & 11.04 & 98.79 & $-1.43\pm0.03$ & $8.25\pm0.18$ & $-8.716\pm0.009$ & $-9.9\pm0.2$ & $1.1\pm0.2$ & E \\
    \textbf{dw1408$+$56} & 10.36 & 58.48 & $-0.73{+0.04\atop-0.03}$ & $8.41\pm0.18$  & $-9.92\pm0.02$ & $-10.0\pm0.2$ & $0.1\pm0.2$ & E \\
    \textbf{UGC 4527} & 10.21 & 52.03 & $-1.58{+0.11\atop-0.08}$ & $8.27\pm0.18$ & $-10.18\pm0.01$ & $-10.1\pm0.2$ & $-0.1\pm0.2$ & B 
    \enddata
    \tablenotetext{a}{The stellar metallicity is reported based on the \texttt{Bagpipes} posterior. Note that we used a flat prior defined from $-2 \leq \log Z/Z_\odot \leq 0$, so especially low-metallicity galaxies may be strongly constrained by the lower bound we adopted on the prior. Additionally, the two galaxies observed with Keck/LRIS -- UGCA 285 and LeG 26 -- are found to be best fit by more metal-rich, near $Z = Z_\odot$, models; given flux calibration concerns for these two galaxies, it is possible that the metallicity of the stellar population is inferred erroneously.}
    \tablenotetext{b}{The summarized star formation history designated here is from the qualitative character of the SFH  returned by \texttt{Bagpipes} (see \S \ref{subsubsec:sed} and Appendix \ref{app:spec}): \textit{sustained} (S),  \textit{bursty} (B), \textit{quiescent} (Q), and \textit{episodic} (E).}
    \tablenotetext{c}{Gas-phase metallicity determined from the $O_{32}$ strong line relation presented in \citet[][]{Curti.etal.2017} as the $N_2$ line ratio is out of range due to H$\alpha$ being a near non-detection.}
\end{deluxetable*}

We use the \texttt{Bagpipes} \citep{Carnall.etal.2018, Carnall.etal.2019b} SED fitting code to estimate the metallicity and star formation history (SFH) from the observed optical spectra. To consistently fit SFHs for a wide range of galaxies, like those in the QUEST Dwarfs sample, we select the \citet{Iyer.etal.2019} non-parametric SFH model, which is an improvement on the Dense Basis SFH fitting introduced in \citet{Iyer.Gawiser.2017}. We allow \texttt{Bagpipes} to fit for the SFH in seven time bins with a Dirichlet prior \citep{Leja.etal.2017, Leja.etal.2019}. Based on the tests in \citet{Iyer.etal.2019}, we adopt $\alpha = 5$, where $\alpha$ is the Dirichlet prior parameter that defines the correlation between the resulting time bins, essentially tuning the smoothness or stochasticity of the star formation history. 

All spectra are given the same priors fixed at their measured redshift\footnote{Where we rely on \texttt{Bagpipes} to recover the redshift, we use a flat prior of $0 \leq z \leq 0.005$. This is the case for UGCA 285, UGC 7485, and UGC 4527, where the SNR and/or wavelength coverage is not sufficient below 5000 \AA~to directly measure the galaxies' $cz$ from Balmer absorption lines.} and are fit with the \texttt{MultiNest} \citep{Feroz.Hobson.2008, Feroz.Hobson.Bridges.2009, Feroz.Skilling.2013} algorithm implemented via \texttt{PyMultiNest} \citep{Buchner.etal.2014}. We adopt flat priors for velocity dispersion ($1 - 1000$ km s$^{-1}$) SFR ($10^{-6} - 1\, \msun \, \mathrm{yr}^{-1}$), stellar metallicity ($10^{-2} - 1\, Z_\odot$), and stellar mass formed ($10^4 - 10^{10}\, \msun$), with the latter three flat priors defined in $\log_{10}$. 

We include a nebular emission component in the spectral fit, but fix the ionization parameter to $U = 10^{-3}$, which is both standard and consistent with the value \texttt{Bagpipes} returned in tests where we allowed this value to vary. Additionally, following \citet{Carnall.etal.2019b}, we add a white noise component to the spectrum, which functionally rescales the spectroscopic errors and results in somewhat more ``realistic'' confidence intervals associated with the posterior distributions from the fit. We directly mask the atmospheric A- (7600-7630 \AA) and B-bands (6860-6890 \AA) in the input spectra to mitigate the impact of their absorption on the spectral fits.
We do not include a dust model in the spectral model as the QUEST Dwarfs are low-mass and relatively low-metallicity, which suggests that they should also be dust-poor \citep[e.g.,][]{Fisher.etal.2014,DeVis.etal.2019,Li.etal.2019}.

The best-fit stellar metallicity of each galaxy is presented in column (4) of Table \ref{tab:derived}, and we characterize the extended SFH returned by \texttt{Bagpipes} (see Appendix \ref{app:spec}) in column (9) of Table \ref{tab:derived}. We divide those returned SFHs into four categories: \textit{sustained}, where star formation is roughly constant over the galaxy's lifetime; \textit{bursty}, where there is a clear gap between episodes of star formation; \textit{quiescent}, where the recent star formation rate has fallen off precipitously and the specific star formation rate is now below $10^{-11} \, \mathrm{yr^{-1}}$; and \textit{episodic}, where the star formation history is more complex and does not fit neatly into one of the other categories.
Given the extensive available data, we infer stellar mass from broadband imaging which captures the total star light and is therefore not subject to slit losses (see \S\  \ref{sec:mass}), and, where available, we additionally constrain the recent SFH from \textit{GALEX} photometry and emission line fitting (Appendix \ref{app:photspec}); as such, no spectrophotometric or aperture correction is applied when fitting the data with \texttt{Bagpipes}.
The optical spectra and their \texttt{Bagpipes} fits are shown in Appendix \ref{app:spec}. 

\subsection{Inferring Mass}
\label{sec:mass} 

We adopt the distances derived in \S\,\ref{sec:distance} to convert the galaxies' integrated magnitudes (see \S\,\ref{sec:photometry} and Table \ref{tab:phot}) to absolute magnitudes and apply the color-based photometric stellar mass calibration for low-mass galaxies presented in \citet[][their equation 13]{delosReyes.etal.2025}. This relation is intended to minimize mass-dependent systematic over- and under-prediction of galaxy stellar mass, which can otherwise bias inferred masses by up to $\sim$half a dex.  Note that the QUEST Dwarfs' stellar masses are reported with an uncertainty of 0.2 dex on top of the photometric and distance uncertainties, which encompasses the intrinsic scatter of the \citet[][]{delosReyes.etal.2025} relation and is consistent with the typical systematic error on color-based stellar mass estimates.

Independent stellar continuum is not detected for LEDA 1722851-em, the  H$\alpha$-bright region associated with LEDA 1722581. Instead, we report here the \texttt{Bagpipes} inferred mass of $2.3 \times 10^8 \, \msun$ for that companion object, which is just a factor of $\lesssim3$ lower than the stellar mass inferred for the main body of LEDA 1722581 by \texttt{Bagpipes}. For the main body of LEDA 1722581, \mstar\ estimated from integrated MaNGA spectroscopy using \texttt{Bagpipes} is consistent with \mstar\ inferred from photometry ($\log\,\mstar/\msun = 8.8$).  The results are presented in column (6) of Table \ref{tab:dat}.

To infer the halo mass of these galaxies, we use the dwarf galaxy stellar mass-halo mass relation (SHMR) from \citet[][]{Manwadkar.Kravtsov.2022}, which is consistent with both \citet[][]{Read.etal.2017} and \citet[][]{Nadler.etal.2020} for dwarf galaxies and can also be reliably extrapolated to higher stellar and halo masses, where it is consistent with e.g., the \citet{BehrooziWechslerConroy.2013} SHMR. The virial radius, $R_{\rm vir}$, which we define here as $R_{200c}$, is then computed for each galaxy, assuming $H_0 = 70\;\mathrm{km~s^{-1}~Mpc^{-1}}$.  The inferred halo mass and halo radius of each galaxy are presented in columns (2) and (3) of Table \ref{tab:derived}.

\begin{figure*}
    \centering
    \includegraphics[width = \linewidth]{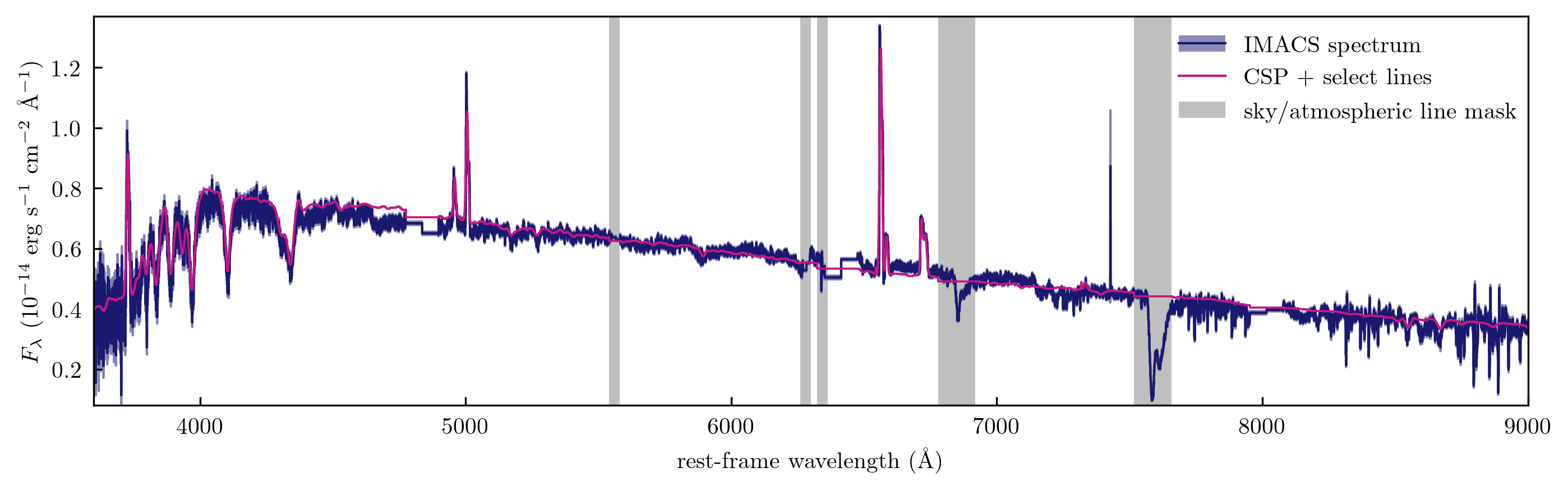}
    \caption{The optical spectrum of UGC 75 (dark blue), overplotted by the CSP model with the nebular lines we fit added (pink). Note that for illustration purposes the normalization of the CSP model is set \textit{globally} here with the line properties adjusted accordingly.
    \label{fig:fullspecfit}}
\end{figure*}

\subsection{Deriving Star Formation Histories} \label{sec:sf}

In addition to the star formation history derived from the \texttt{Bagpipes} SED fitting (see Appendix \ref{app:spec})
we measure the (specific) SFR at both $\sim100$ Myr and $\sim 10$ Myr with FUV photometry and the H$\alpha$ line respectively. We use the integrated FUV photometry (\S\,\ref{sec:photometry}) and our distance estimates (\S\,\ref{sec:distance}) to infer the galaxy's global SFR on $\sim 100$ Myr timescales using equation 3 of \citet[][]{Lee.etal.2009a}. We then convert this SFR to a specific SFR, sSFR, normalizing by the stellar mass from \S\,\ref{sec:mass}. The shorter timescale sSFR is derived from the equivalent width of the H$\alpha$ line as described in \S\,\ref{subsubsec:lines} below. These specifically measured star formation rates are compared against the \texttt{Bagpipes} SFHs in Figures \ref{fig:specsfhb} through \ref{fig:specsfhe}.

The ratio between integrated H$\alpha$ and FUV fluxes has been used as a proxy for the burstiness of dwarfs' star formation histories \citep[e.g.,][]{Lee.etal.2009a,Weisz.etal.2012,Perry.etal.2025}. As we do not have narrow-band H$\alpha$ imaging for the QUEST Dwarfs, we instead adopt $\log(\mathrm{sSFR(H\alpha)/sSFR(FUV)})$ here, which should be proportional to the flux ratio\footnote{Though the sSFR is not linear with respect to the H$\alpha$ equivalent width in \citealt{Belfiore.etal.2018}, that non-linearity is attributed to the stellar mass correction, in-keeping with other relations where $\mathrm{SFR\sim F(H\alpha)}$.}. Critically, the sSFRs we compare are not computed over the same area -- the FUV-inferred sSFR is based on independent estimates of SFR from integrated \textit{GALEX} imaging and $\mstar$, while the H$\alpha$-inferred sSFR is estimated from the spectral line -- which introduces an additional spatial explanation for discrepancies between the two sSFRs independent of temporal burstiness.  We report H$\alpha$-inferred and FUV-based sSFRs, as well as the ratio of the two quantities in columns (6)--(8) of Table \ref{tab:derived}, respectively.

\begin{deluxetable*}{l|ccccccc}
\tablewidth{\linewidth}
    \tablecaption{Measured absorption features. \label{tab:ion}}
    \tablehead{\textbf{Galaxy}  & \boldmath\textbf{$d_\mathrm{proj}/R_\mathrm{vir}$}\unboldmath & \boldmath\textbf{$\log N_\mathrm{\textsc{Hi}}/ \mathrm{cm}^{-2}$}\unboldmath &
    \boldmath\textbf{$\log N_\mathrm{Si\textsc{ii}}/ \mathrm{cm}^{-2}$}\unboldmath & \boldmath\textbf{$\log N_\mathrm{\textsc{Cii}}/ \mathrm{cm}^{-2}$}\unboldmath & \boldmath\textbf{$\log N_\mathrm{Si\textsc{iii}}/ \mathrm{cm}^{-2}$}\unboldmath &
    \boldmath\textbf{$\log N_\mathrm{Si\textsc{iv}}/ \mathrm{cm}^{-2}$}\unboldmath & \boldmath\textbf{$\log N_\mathrm{C\textsc{iv}}/ \mathrm{cm}^{-2}$}\unboldmath}
    \startdata
    \textbf{PGC 4143} & 0.570 & ... & ... & ... & ... & $...$ & $13.42{+0.07\atop-0.08}$ \\
    \textbf{} & 0.607 & $14.7\pm0.1$ & $<12.89$ & $13.4\pm0.2$ & ... & $12.975{+0.07\atop-0.075}$ & $13.3\pm0.1$ \\
    \textbf{} & 0.629 & $14.21{+0.08\atop-0.06}$ & $<11.92$ & $<13.05$ & ... & ... & $<12.875$ \\
    \textbf{UGCA 285} & 0.061 & $>14.37$& $<11.85$ & $<12.73$ & ... & $<12.83$ & $13.709{+0.07\atop-0.085}$ \\
    \textbf{UGC 7370} & 0.767 & $14.42{+0.03\atop-0.02}$& $<11.96$ & $<12.60$ & ... & $<12.37$ & $13.78{+0.05\atop-0.06}$ \\
    \textbf{PGC 41458} & 0.251 & $13.58{+0.08\atop-0.07}$& $<12.00$ & $<12.985$ & $<11.90$ & $<12.12$ & ... \\
    \textbf{LeG 26} & 0.161 & $>14.56$ 
    & $13.3\pm0.2$ & $14.03{+0.08\atop-0.09}$ & $13.19{+0.06\atop-0.065}$ & $13.16\pm0.08$ & $14.24{+0.12\atop-0.08}$ \\
    \textbf{UGC 75} & 0.949 & $13.55{+0.05\atop-0.06}$ & $<12.02$ & ... & $<11.52$ & $<12.06$ & $<12.24$ \\
    \textbf{UGC 9126} & 0.180& $19.81\pm0.01$ & $14.1{+0.2\atop-0.1}$ & $>14.64$ & ... & $13.34\pm0.04$ & ... \\
    \textbf{UGC 7485} & 1.192 & $13.22{+0.07\atop-0.09}$ & $<12.15$ & $<12.51$ & $<11.77$ & $<11.97$ & $<12.30$ \\
    \textbf{LEDA 1722581} & 0.178 & $>15.56$ & $14.40 \pm 0.04$ & $>15.58$ & $>14.12$ & $14.13\pm0.02$ & $15.1{+0.5\atop-0.1}$\\
    \textbf{PGC 1818175} & 0.179 & $>14.84$ & $<12.46$ & $13.8{+0.1\atop-0.2}$ & $<12.75$ & $<12.90$ & ...\\
    \textbf{UGC 5427} & 1.316 & $13.90{+0.045\atop-0.04}$ & $<11.61$ & $<12.87$ & $<11.63$ & $<11.94$ & $<12.92$\\
    \textbf{NGC 5486} & 0.119& $>15.14$ & $15.0\pm0.3$ & $>14.99$ & $>13.82$ & $13.67\pm0.03$ & ... \\
    \textbf{dw1408+56} & 0.441 & $>13.81$ & $<11.69$ & $<13.51$ & ... & $12.8\pm0.1$ & ... \\
    \textbf{UGC 4527} & 0.139 & $>14.76$& $12.4\pm0.1$ & $13.47\pm0.05$ & $13.16{+0.04\atop-0.03}$ & $12.85\pm0.03$ & $13.58{+0.04\atop0.05}$
    \enddata
\end{deluxetable*}

\subsection{Measurements of Spectral Features} \label{subsubsec:lines}

We measure the equivalent widths (EWs) and fluxes of a suite of optical nebular lines, including [\textsc{Oii}]$_{3728}$, H$\beta$, [\textsc{Oiii}]$_{4960}$, [\textsc{Oiii}]$_{5008}$, H$\alpha$, and [\textsc{Nii}]$_{6585}$. For each galaxy, we use a low-order polynomial to approximate the local continuum and use stellar phase metallicity and non-parametric star formation history from the \texttt{Bagpipes} fitting to create complex stellar population (CSP) models of only stellar continuum and absorption with \texttt{FSPS} against which lines can be fit. To account for potentially imperfect flux calibration across the entire range of the fit, the normalization of the CSP component is re-evaluated in each spectral fitting region. Measurements of line fluxes and equivalent widths are presented in Table \ref{tab:lines} and an example of the results of this fitting is shown in Figure \ref{fig:fullspecfit}.

We infer gas-phase metallicities for the QUEST Dwarfs based on their observed $N_2$ index with $N_2\equiv\log\,{\rm N\textsc{ii}}/{\rm H\alpha}$, following \citep[][see also \citealt{Kewley.Nicholls.Sutherland.2019} for a review]{Pettini.Pagel.2004}.  The results are presented in column (5) of Table \ref{tab:derived}.  From the equivalent width of H$\alpha$, ${\rm EW}({\rm H\alpha})$, we infer sSFR using the empirical, MaNGA-derived EW(H$\alpha$)-sSFR relation from \citet[][]{Belfiore.etal.2018}. This sSFR is then independent of any slit- or fiber-loss and agnostic to each galaxy's estimated stellar mass (column 6 of Table \ref{tab:derived})

The strength of the 4000\AA~break ($D_n4000$) can be used as a diagnostic for the age of a galaxy's stellar population \citep{Bruzual:1983, Hamilton.1985}. However, the age-dependence of $D_n4000$ becomes weaker with decreasing stellar metallicity \citep[e.g.,][]{vandenBergh.1963, Paulino-Afonoso.etal.2020}. We measure $D_n4000$ here, adopting the red and blue continuum definitions from \citet[][]{Balogh:1999}. These measurements are presented in the rightmost column of Table \ref{tab:lines}.

\subsection{Environment}
\label{subsec:environ}

Using the galaxy distance and mass derived in \S\S\,\ref{sec:distance} and \ref{sec:mass}, we examine the ``neighborhood'' around the QUEST Dwarfs. We show these galaxies in the context of their immediate environments ($\pm 2$ deg) using the 50 MGC \citep{Ohlson.etal.2024} and the DESI DR1 \citep{DESI.DR1} catalog to find galaxies within $300\,\mathrm{km~s^{-1}}$ of the target dwarf galaxy's measured radial velocity. For these non-target sources, we use the ``best'' stellar mass from the 50 MGC where available, or, for galaxies that are only in the DESI catalog, we compute the stellar mass using catalog photometry\footnote{Though ``fragmentation'' or ``shredding'' remains a problem for automated photometry of nearby galaxies in large surveys, we adopt the DESI catalog values as any correction for this effect would be beyond the scope of our analysis. The determination of the environments of QUEST Dwarfs is only minimally impacted by the stellar mass of non-target galaxies; instead, given their projected positions, even significantly revised non-target galaxy mass estimates would not affect the environment classification here.} and the \citet[][]{delosReyes.etal.2025} relation. We follow \S\,\ref{sec:mass} and use the \citet[][]{Manwadkar.Kravtsov.2022} SHMR to infer halo masses of these neighboring galaxies and compute their $R_{\rm vir}$ based on $R_{200c}$ for $H_0 = 70\;\mathrm{km~s^{-1}~Mpc^{-1}}$.

We show the QUEST Dwarfs' environments in Figure \ref{fig:env}. Galaxies are placed into one of three categories based on their environments. \textit{Isolated} galaxies are $\gtrsim 2 R_\mathrm{vir}$ from their nearest neighbor. \textit{Group} galaxies are in close associations with other galaxies or are part of known groups. Galaxies are categorized as \textit{interacting} if their virial radius overlaps significantly with the virial radius of a close neighbor (in both projected distance and velocity). Note that there is significant variety in the environment of \textit{group} and \textit{interacting} systems.

{In such complex environments, a fundamental challenge is determining which galaxy is physically associated with an observed absorber. In an extensive study, \citet{Qu:2023} showed that CGM absorption properties in group environments are primarily driven by the galaxy closest to the QSO sightline, rather than by the most massive galaxy or by mass-weighted group properties. Guided by this finding, we associate each absorber with the galaxy closest to the QSO sightline in the analysis below.}

\begin{figure}
    \centering
    \includegraphics[width = \linewidth]{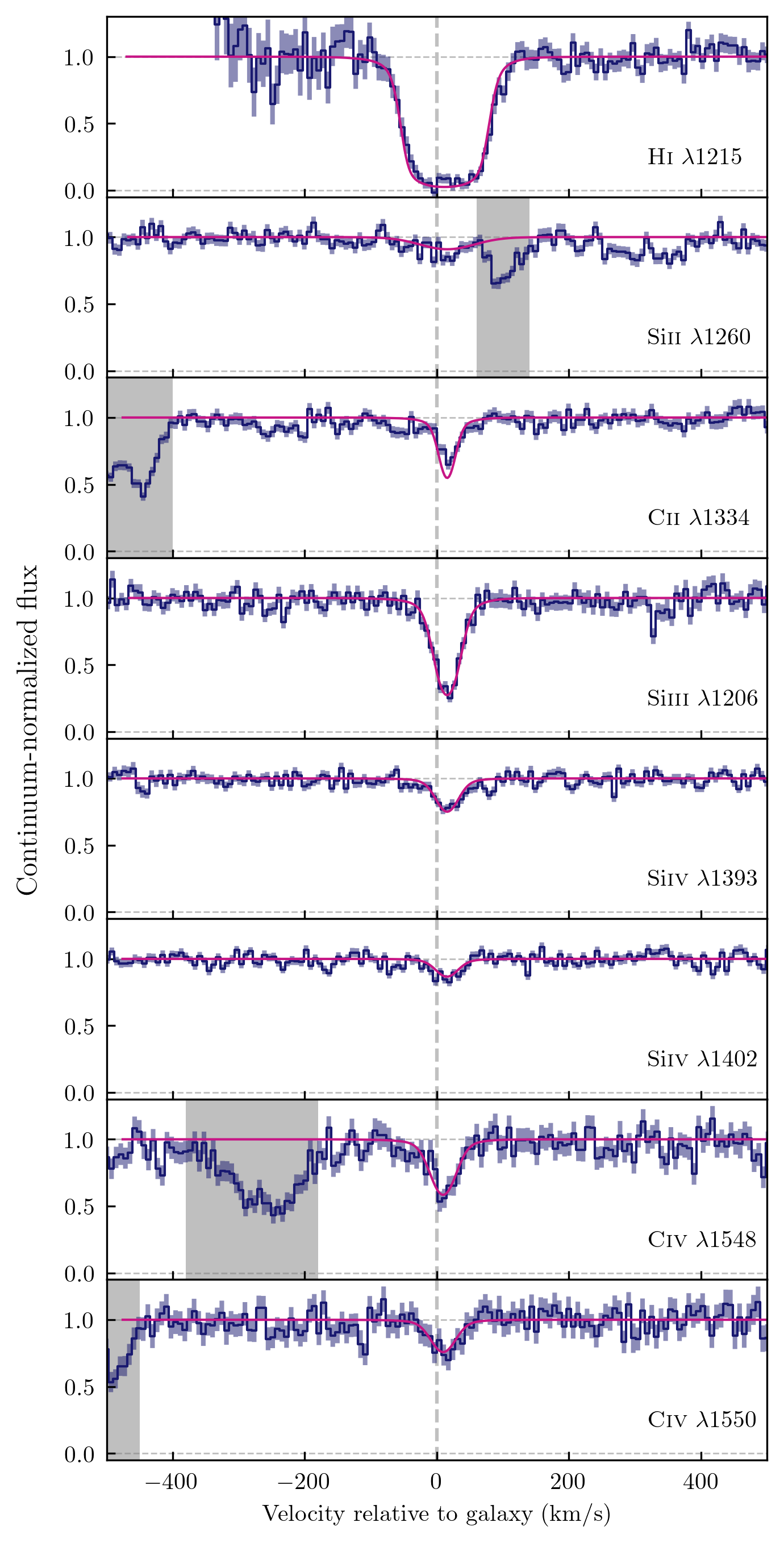}
    \caption{A suite of absorption transitions identified at $d_{\rm proj}\approx 7$ kpc (or $\approx 0.14\,R_{\rm 200}$) from UGC4527, a dwarf galaxy of $\log\,\mstar/\msun\approx 7.6$ at $z=0.00242$ (or $cz\approx 725$ \kms; see Table \ref{tab:dat}).  Zero velocity corresponds to the systemic redshift of the galaxy (see \S\ \ref{subsec:cz} and Table \ref{tab:dat}). In addition to \ion{H}{1} at the top, we detect moderately strong ionic transitions due to \ion{Si}{2}, \ion{C}{2}, \ion{Si}{3}, \ion{Si}{4}, and \ion{C}{4} (from top to bottom). Gray regions demarcate absorption not associated with the relevant line, while best-fit Voigt profile models are overplotted in pink.
    \label{fig:dwarf_cgm}}
\end{figure}

\section{Properties of the CGM around QUEST Dwarfs} \label{sec:cgmanalysis}

The core aim of the QUEST Dwarfs sample is to connect properties of the CGM, derived from absorption spectroscopy of FUV-bright background quasars, to the galaxies' ISM and stellar populations. To determine the CGM properties of QUEST Dwarfs, we perform an absorption-line search in the FUV spectra of the background QSOs described in \S\,\ref{subsec:fuvspec}.  
Specifically, we search for a suite of strong ionic lines from Si, C, and O as well as hydrogen near the galaxies' central radial velocities.  Prominent FUV absorption transitions of the diffuse CGM covered by \textit{HST}/COS include \ion{H}{1} Lyman series, \ion{C}{2}\,1334, \ion{C}{4}\,1548, 1550, \ion{Si}{2}\,1260, 1304, 1526, \ion{Si}{3}\,1206, and \ion{Si}{4}\,1392, 1402 (see Figure \ref{fig:dwarf_cgm} for an example and Figures \ref{fig:abs1}, \ref{fig:abs2}, and \ref{fig:abs3} for the full sample). We perform a Voigt profile analysis of all absorption features following the steps described in \citet{Qu:2026} and adopting the minimum number of components necessary to capture the observed asymmetries of the absorption profiles.

Along all but two of the QUEST Dwarfs sightlines, \ion{H}{1} \lya\ absorption is detected at the redshifts of the dwarfs.  For LeG 26, the \lya\ absorption is contaminated by strong and complex absorption features due to \ion{Ne}{4}\,543 originating in the high-speed outflows from the background QSO, 2MASS J10512569$+$1247462.  In addition, the sightline toward LBQS\,0107$-$0232 around PGC\,4143 has no COS G130M data available for placing constraints at wavelengths below 1450 \AA\ (see Table \ref{tab:qsos}).  Incidentally, this dwarf has two additional sightlines probing its CGM at two other locations, and both exhibit a strong \lya\ absorption feature.  Because \lya\ is a strong transition and, in seven cases, saturated, and the available \textit{FUSE} spectra are not of sufficiently high quality to constrain $N({\rm HI})$ based on the flux discontinuity at the Lyman limit, we are only able to place
a 2-$\sigma$ lower limit to $N({\rm HI})$ for these saturated \lya\ absorbers. For the remaining eight sightlines through seven dwarf halos, we can determine $N({\rm HI})$, and its associated 1-$\sigma$ uncertainties based on the Voigt profile analysis. For each \ion{H}{1} absorber, we search for associated ionic transitions and perform a Voigt profile analysis, using the consistent line centroid of the \ion{H}{1} line.  In cases when the ionic transitions are not detected, we report a 2-$\sigma$ upper limit to the underlying ionic column densities for a fixed Doppler velocity width of $15$ \kms\ \citep[typical of the line widths observed in low-ionization transitions; see e.g.,][]{Zahedy:2019, Zahedy:2021, Qu:2022}, using a Bayesian framework described in \citet[][see also \citealt{Zahedy:2021}]{Qu:2022}. 
We report all available column density measurements and their associated uncertainties in each dwarf galaxy halo in Table \ref{tab:ion}. The absorption profiles for different absorption species detected in the vicinity of UGC4527 are presented in Figure \ref{fig:dwarf_cgm} for illustration.  The remaining sightlines are presented in the Appendix for completeness. 

Note that \citet[][]{Stocke.etal.2013} previously reported an \ion{H}{1} column density for absorption associated with UGC 4527, consistent with our lower limit due to line saturation, and absorption measurements of \ion{C}{2}, \ion{C}{4}, \ion{Si}{2}, and \ion{Si}{3} associated with UGC\,7485 in the FUV spectrum of QSO 4C21.35 (PG1222$+$216) were previously reported by \citet{Zheng.etal.2024}, with non-detections for both \ion{H}{1} and \ion{Si}{4}. For UGC 7485, we measure the \ion{H}{1} absorption directly and place more stringent upper limits on the remaining ions, \ion{C}{2}, \ion{C}{4}, and \ion{Si}{2}-\ion{Si}{4}. 
Although the data presented here are archival, all other sightlines are newly analyzed for absorption by the target dwarf galaxies.

\begin{figure*}
    \centering
    \includegraphics[width = \textwidth]{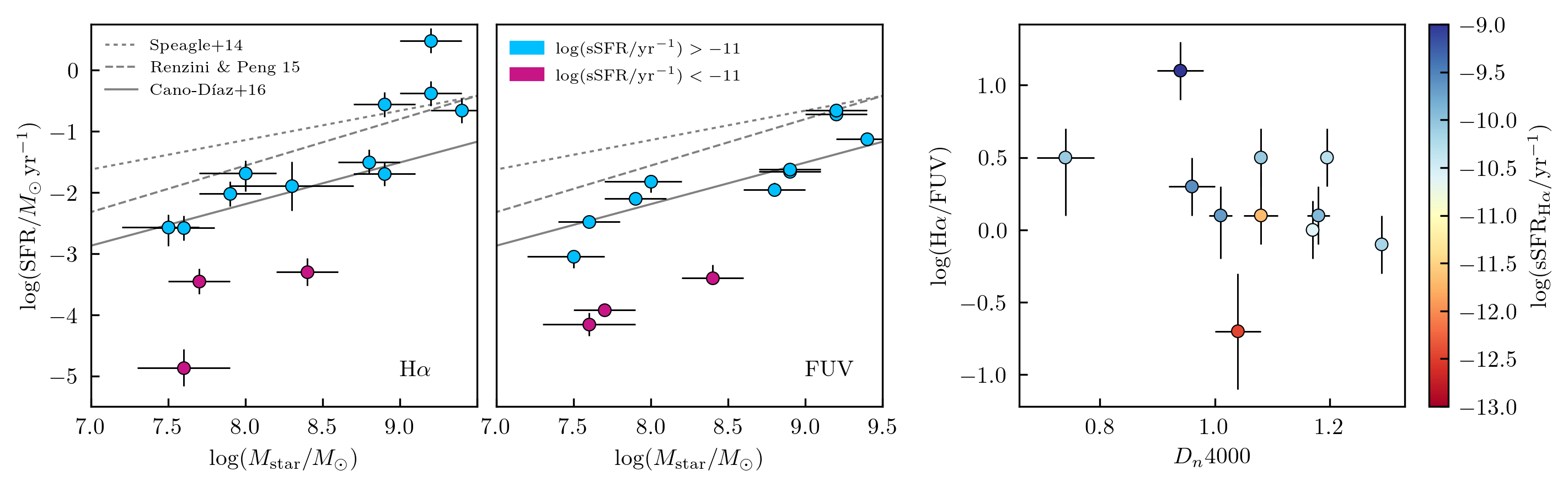}
    \caption{Summary of the star formation histories of the first release QUEST Dwarfs sample.  The best-fit $z = 0$ star-forming sequence from \citet[][]{Speagle.etal.2014}, \citet[][]{Renzini.Peng.2015}, and \citet[][]{Cano-Diaz.etal.2016} compared to the presented sample of galaxies. H$\alpha$-derived SFRs are shown at left, and FUV SFRs are shown in the central panel. Galaxies are colored according to whether they are star-forming (blue) or quiescent (pink). The recent star formation history of each galaxy, as described by $\log(\mathrm{H}\alpha/\mathrm{FUV})$, compared to the measured 4000\AA~line break is shown at right. As expected, more quiescent with with $\log\,{\rm sSFR/yr^{-1}}<-11$ exhibit a significant 4000-\AA\ break. However, little correlation is found between $D_n4000$ and individual SFR indicators.   
    \label{fig:sfms}}
\end{figure*}

\section{Discussion} \label{sec:discussion}

A primary goal of the QUEST Dwarfs program is to establish a dataset that enables comprehensive studies of the baryon cycle in low-mass galaxies, from star formation and feedback within galaxies to the interplay between their interstellar media, stellar populations, and gaseous halos.  As an initial step, we have constructed the first release QUEST Dwarfs sample of 14 galaxies, measured their spectral and photometric properties, and placed sensitive constraints on their CGM based on available archival FUV spectra of the background QSOs.  Here we first place the QUEST Dwarfs sample in the broader context of the larger galaxy population by examining fundamental scaling relations, including the stellar and gas-phase mass–metallicity relations (MZRs) and the star-forming main sequence \citep[e.g.,][]{Noeske.etal.2007}, to assess whether the sample exhibits the expected trends observed in nearby galaxies. We then continue to investigate the connections between galaxy properties and CGM absorption properties.

\subsection{Star formation and chemical enrichment in QUEST Dwarfs}
\label{sec:sfh}

We first examine the star formation history of the first release QUEST Dwarfs by comparing the inferred SFR and stellar mass in Figure \ref{fig:sfms} (two left panels).  In comparison to published relations for the local galaxy populations from \cite{Speagle.etal.2014}, \cite{Renzini.Peng.2015}, and \cite{Cano-Diaz.etal.2016}, 11 of the 14 galaxies in this first-release sample are consistent with expectations from the star-forming main sequence, while three are qualified quiescent galaxies with $\log\,{\rm sSFR/yr^{-1}}<-11$.  While FUV-inferred SFRs, which are expected to reflect the mean SFR on time scales of 100 Myr or longer, tend to be systematically lower\footnote{Both H$\alpha$ and the FUV star formation rates may need to be revised downward to account for sub-solar [Fe/H], with rough estimates suggesting that for a galaxy with $\mathrm{[Fe/H]} = -1$, FUV SFRs are overestimated by $0.07$ dex and H$\alpha$ SFRs are overestimated by 0.4 dex \citep[][]{Kennicutt.Evans.2012}. Given uncertainties on these corrections, we do not adopt them here; however, corrections of that nature alleviate some of the discrepancy between the two SFR indicators, while making tighter the correlation between $\log\,\mathrm{H}\alpha/\mathrm{FUV}$ and $D_n4000$.} than the H$\alpha$-inferred SFRs, the division between quiescent and star-forming in the QUEST Dwarfs sample remains clear.
  
\begin{figure*}
    \centering
    \includegraphics[width = \linewidth]{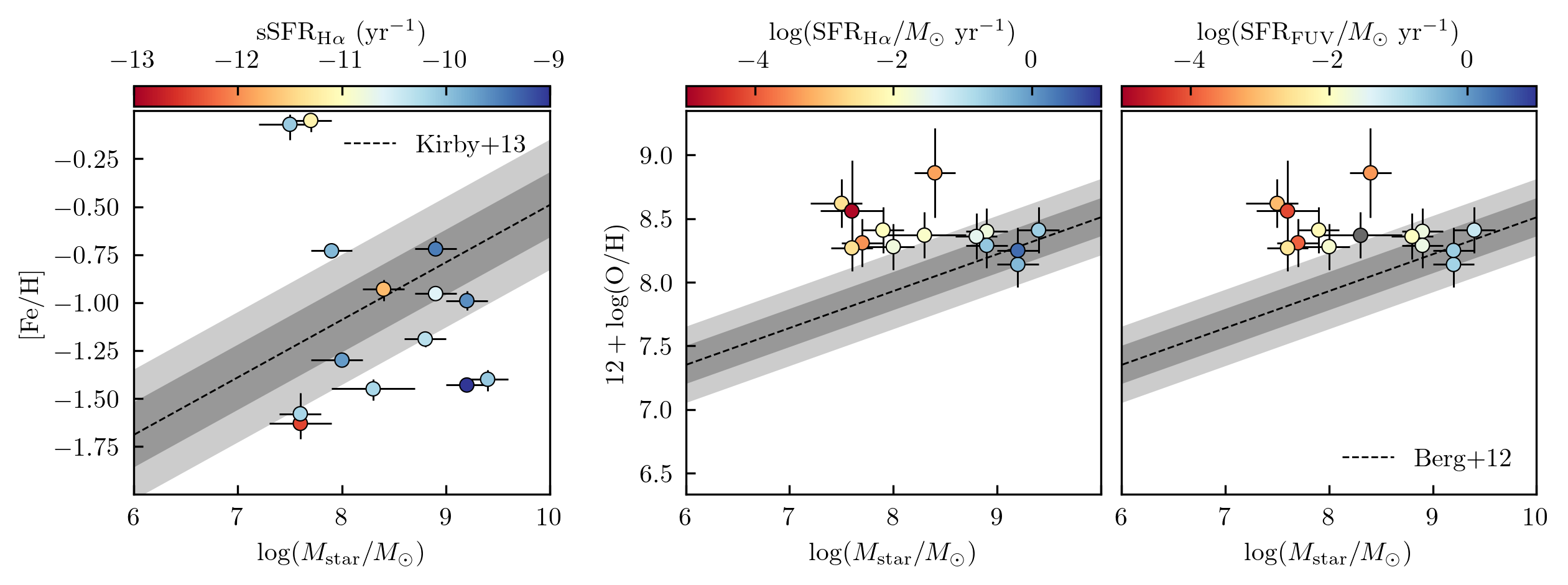}
    \caption{The low-mass stellar \citep[][left]{Kirby.etal.2013} and gas-phase \citep[][middle and right]{Berg.etal.2012} mass-metallicity relation compared to the metallicities reported for the first release subsample of QUEST Dwarf galaxies, at \textit{left} colored by the specific star formation rate over the last $\sim10$ Myr, \textit{middle} colored by SFR over the last $\sim10$ Myr, and \textit{right} colored by SFR over the last $\sim100$ Myr. Note that the two very high [Fe/H]$\sim 0$ galaxies only have Keck/LRIS spectra available, for which the calibrations on the blue side are uncertain. 
    As \texttt{Bagpipes} takes the shape of the continuum into account in fitting, those metallicities are likely more uncertain than is expressed by the error bar. Gas-phase metallicity scales inversely with star formation rate.
    \label{fig:mzr}}
\end{figure*}

We also compare $\log(\mathrm{H}\alpha/\mathrm{FUV})$ and $D_n4000$ in the right panel of Figure \ref{fig:sfms}.  The ratio between H$\alpha$ and FUV continuum is adopted as a proxy for tracking the recent star formation history of the galaxies. At the same time, $D_n4000$ is often used as an indicator of stellar population age \citep[e.g.,][]{Bruzual:1983}, although it is also sensitive to stellar metallicity given that the flux discontinuity is a result of line blanketing of a large number of metal lines \citep[e.g.,][]{Balogh:1999}.  Notably, there is little correlation between the $D_n4000$ and individual SFR indicators for these galaxies, 
nor with $\log\,\mathrm{H}\alpha/\mathrm{FUV}$ (see Tables  \ref{tab:derived} and \ref{tab:lines}, and Figure \ref{fig:sfms}).  Such a lack of correlation between $D_n4000$ and the star-formation indicators may reflect the combined effects of age and metallicity, as $D_n4000$ is sensitive to both stellar population age and metal-line blanketing \citep[e.g.,][]{Balogh:1999}. The broad range of stellar metallicities in the QUEST Dwarfs sample ($-2 \lesssim {\rm [Fe/H]} \lesssim 0$, see Table \ref{tab:derived}) may therefore contribute to the observed scatter. 

In Figure \ref{fig:mzr}, we compare the stellar metallicity, [Fe/H], fit by \texttt{Bagpipes} and ISM gas-phase metallicity $12+\log(\mathrm{O/H})$ inferred from the $N_2$ line ratio following \citet[][]{Pettini.Pagel.2004} to well-established mass-metallicity relations \citep[][]{Berg.etal.2012, Kirby.etal.2013}.  Considering the stellar and ISM gas-phase MZRs separately is necessary as they exhibit different slopes in this low-mass regime.  The stellar MZR in the QUEST Dwarfs sample follows the declining trend observed in nearby field dwarfs \citep[e.g.,][]{Kirby.etal.2013}, with the exception of UGCA\,285 and LeG\,26. Both galaxies' analysis was done with archival LRIS data that have uncertain blue flux calibrations, resulting in poorly constrained continuum shapes below $\sim\,5000$ \AA. Combined with degeneracies between stellar metallicity and star formation history in the \texttt{Bagpipes} fits, this likely contributes to the observed metallicity offsets of $\lesssim\,0.3$ dex. 

The inferred gas-phase metallicities measured for star-forming galaxies in our sample show a flat distribution with an apparent floor at $12+\log(\mathrm{O/H}) \approx 8.0$, although they are consistent with the \citet{Berg.etal.2012} MZR given relatively large measurement uncertainties.  The constant gas-phase metallicity over the stellar mass range in our sample may be a result of applying integrated nebular line fluxes across the full extent of the galaxies (see \S\,\ref{sec:data}). In the presence of metallicity variations or metallicity gradients \citep[e.g.,][]{Sanchez:2014, Li:2025} within individual galaxies, the integrated ion line fluxes could be influenced by the most metal-enriched regions with the strongest [\ion{N}{2}] lines.  We will explore such potential bias in follow-up spatially resolved studies (e.g., Bardon Soto et al., 2026, in preparation). 
For quiescent galaxies, our gas-phase inferred metallicities are more likely upper limits, even when measurements of the $N_2$ ratio can be made. 
At the same time, the N/O ratio may flatten at low metallicities \citep[e.g.,][]{Strom.etal.2018}, though the regime in which this seems to occur is $\gtrsim0.2$ dex below our galaxies' apparent $12 + \log(\mathrm{O/H})$ floor.

In summary, the QUEST Dwarfs sample is broadly representative of low-metallicity field dwarf galaxies. Most galaxies in the sample lie on the star-forming main sequence, while three are classified as quiescent based on their low specific star formation rates. The stellar and gas-phase metallicities are generally consistent with established mass–metallicity relations for nearby dwarf galaxies, although the current sample size and measurement uncertainties limit a more detailed assessment of intrinsic scatter and secondary trends.

\subsection{The correlation between CGM properties and galaxy properties}
\label{sec:scaling}

Having established the star formation histories of the QUEST Dwarfs, we next examine the connection between galaxies and the surrounding CGM through the distribution of chemically enriched gas. Silicon is particularly well suited for this purpose because the available FUV spectra encompass the dominant ionization states, \ion{Si}{2}, \ion{Si}{3}, and \ion{Si}{4} \citep[e.g.,][]{Muratov.etal.2017}. The combined column densities of these ions provide a nearly ionization-independent measure of the silicon content of the CGM, allowing us to investigate how the metal budget and its spatial distribution relate to the stellar and ISM properties of the host galaxies \citep[see also][]{Johnson.etal.2017, Zheng.etal.2020, Fox:2026}.

\begin{figure*}
    \centering
    \includegraphics[width = \textwidth]{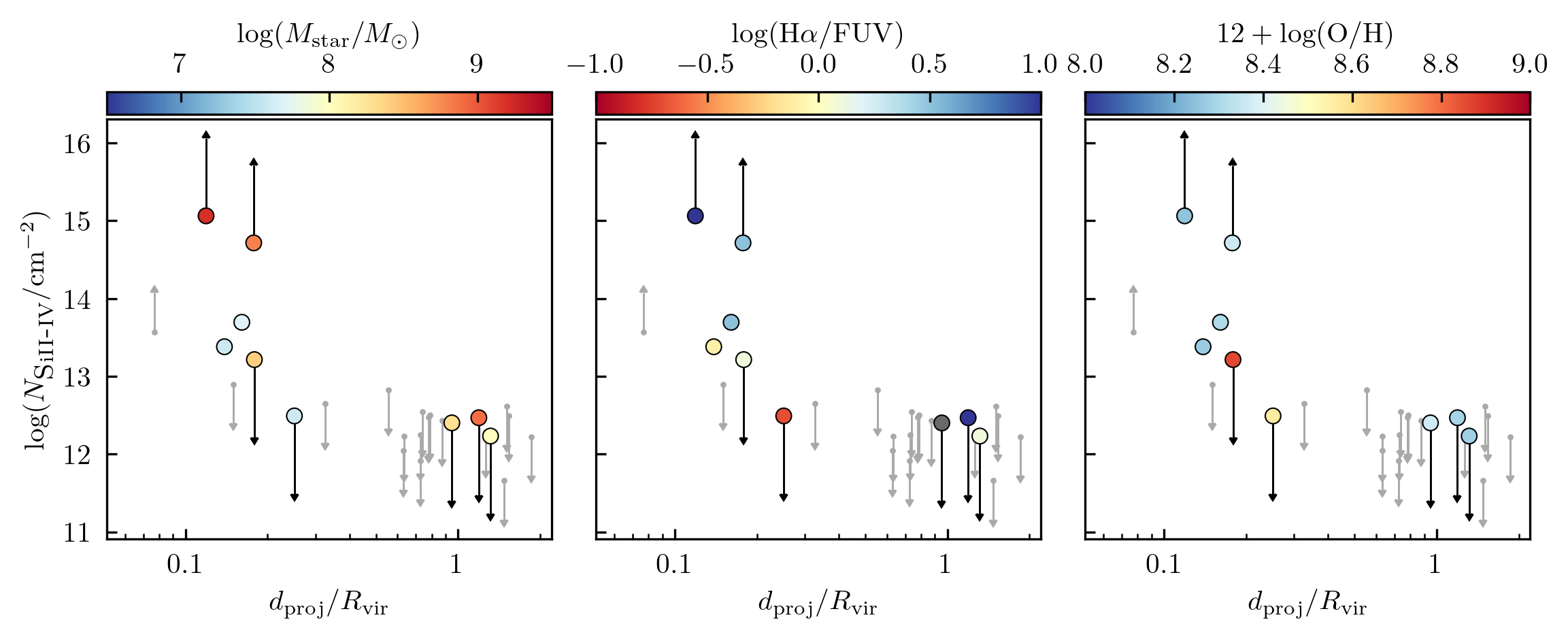}
    \caption{The total column density of Si from sightlines where all of \ion{Si}{2}, \ion{Si}{3}, and \ion{Si}{4} can be measured as a function of $d_\mathrm{proj}/R_{\rm vir}$. We use total Si as a proxy for the overall metal content of the CGM. The panels are colored by stellar mass (\emph{left}), recent star formation history ($\log(\mathrm{H}\alpha/\mathrm{FUV})$; \emph{center}), and gas-phase metallicity ($12 + \log(\mathrm{O/H})$; \emph{right}). Literature values for dwarf galaxies outside of the Local Group \citep[from][]{Bordoloi.etal.2014, Bordoloi.etal.2018, Liang.Chen.2014, Johnson.etal.2017, Johnson.etal.2026, Zheng.etal.2024, Mishra.etal.2024} are shown as light gray points.
    \label{fig:cgmism}}
\end{figure*}

Figure~\ref{fig:cgmism} presents the total silicon column density as a function of halo-radius-normalized projected distance, $d_{\rm proj}/R_\mathrm{vir}$, for the nine sightlines in which constraints on all three dominant silicon ions, \ion{Si}{2}, \ion{Si}{3}, and \ion{Si}{4}, are available. For comparison, we also include dwarf galaxies from previous studies. To explore the physical drivers of the observed spatial distribution, the data are color-coded by stellar mass, recent star formation history ($\log[\mathrm{H}\alpha/\mathrm{FUV}]$), and gas-phase metallicity ($12+\log(\mathrm{O/H})$) in the left, middle, and right panels, respectively.

The inclusion of the QUEST Dwarfs has increased by nearly a factor of three the number of available probes of the inner CGM ($d_{\rm proj}<0.3\,R_{\rm vir}$) in low-mass halos, enabling a more robust characterization of the radial distribution of metals. Consistent with previous studies, the total silicon column density, combining Si$^+$, Si$^{2+}$, and Si$^{3+}$, declines rapidly with increasing halo-radius-normalized projected distance. Four of the ten sightlines within $0.3\,R_{\rm vir}$ exhibit a total silicon column density exceeding $\log N(\mbox{\ion{Si}{2}-\small{IV}})=13.5$, corresponding to a covering fraction of $\kappa_{\rm Si}=0.40_{-0.18}^{+0.20}$, whereas only one of the 16 sightlines at $0.3\le d_{\rm proj}/R_{\rm vir}\le1$ exceeds this threshold, corresponding to $\kappa_{\rm Si}=0.06_{-0.05}^{+0.13}$. In contrast, strong \lya\ absorption remains common throughout the halo. Ten of the 11 sightlines within $0.3\,R_{\rm vir}$ exhibit $N(\mbox{\ion{H}{1}})>10^{14}\,\cmjj$, corresponding to a covering fraction of $\kappa_{\rm HI}=0.91_{-0.18}^{+0.07}$, while 18 of the 24 sightlines at $0.3\le d_{\rm proj}/R_{\rm vir}\le1$ satisfy the same criterion, yielding $\kappa_{\rm HI}=0.75_{-0.12}^{+0.09}$. For a halo gas density profile that declines with radius \citep[e.g.,][]{Zahedy:2019, Qu:2023}, the neutral fraction of hydrogen should also decrease outward. The observed \ion{H}{1} distribution therefore likely underestimates the true extent of the total hydrogen reservoir, implying that the contrast between the radial distributions of metals and gas is even stronger than suggested by the \ion{H}{1} measurements alone. Together, these results indicate that chemically enriched cool gas is preferentially concentrated in the inner CGM, while the diffuse hydrogen reservoir extends well beyond $R_\mathrm{vir}$. 

Superimposed on this radial trend, a clear dependence on stellar mass emerges: at a fixed $d_{\rm proj}/R_\mathrm{vir}$, more massive galaxies tend to exhibit higher silicon column densities, in agreement with expectations from cosmological simulations \citep[e.g.,][]{Schaye.etal.2015, Zheng.etal.2024, Piacitelli.etal.2025}. While previous observational studies have reported mixed evidence regarding the role of stellar mass in regulating CGM metal absorption \citep[e.g.,][]{Zheng.etal.2024}, the first-release QUEST Dwarfs sample substantially improves coverage within $0.3\,R_\mathrm{vir}$, where the mass dependence becomes particularly apparent. Together, these results suggest that both projected distance and galaxy mass play important roles in determining the silicon content of the CGM, with distance governing the overall radial decline and stellar mass setting the normalization of the silicon column density profile.

At the same time, galaxies with elevated $\log(\mathrm{H}\alpha/\mathrm{FUV})$, indicative of recent enhancements in star formation activity, and perhaps also lower gas-phase metallicities appear to exhibit higher silicon column densities. However, because SFR and stellar mass are strongly correlated along the star-forming main sequence (Figure \ref{fig:sfms}), we cannot determine from the current sample whether these trends reflect an independent dependence on recent star formation history or are simply a manifestation of the stellar-mass dependence discussed above.  An additional quantity to consider is the possible azimuthal dependence of metal column density relative to the minor axis of star-forming disks, where conical outflows may elevate the observed metal content \citep[see e.g.,][]{Bordoloi:2011, Peroux:2020}.  However, our sample remains too small for drawing a definitive conclusion.  While the QSO sightline appears to pass close to the minor axis of LEDA\,1722581, one of the two galaxies with the strongest total observed silicon column densities, NGC\,5486 appears to be nearly face-on and a robust measurement of the azimuthal alignment is challenging \citep[see e.g.,][]{Venkat:2025}.

\subsection{Chemical enrichment in the vicinities of low-mass dwarf galaxies}
\label{sec:chemistry}

The QUEST dwarf sample also provides a preliminary estimate of the total Si (and metal) mass residing in the inner halo of dwarf galaxies. In all, there are seven sightlines within $1/3 \, R_\mathrm{vir}$ with measurements or limits on three of silicon ionization stages, \ion{Si}{2}, \ion{Si}{3}, and \ion{Si}{4} (see Figure \ref{fig:cgmism}), including both the QUEST Dwarfs and the existing dwarf galaxy literature sample outside of the Local Group. Because of the correlation between galaxy stellar mass and absorber strength described in \S\,\ref{sec:scaling}, we divide these galaxies into ``low-mass'' ($\log M_\mathrm{star}/\msun < 8$) and ``high-mass'' ($\log M_\mathrm{star}/\msun \geq 8$) populations, with each bin containing three and four sightlines respectively. For the low-mass subsample with a median stellar mass of $\langle\log M_\mathrm{star}/\msun\rangle_{\rm med} = 7.6$ and corresponding $R_\mathrm{vir} = 52.0$ kpc, two of the three sightlines have $N_\mathrm{Si} > 10^{13}\,\mathrm{cm}^{-2}$, corresponding to a covering fraction of $\kappa_{\rm Si} \sim 0.66$ for gas exceeding this $N_\mathrm{Si}$ threshold. We estimate a total enclosed Si mass of $\log M_\mathrm{Si\,II-IV}(<0.3\,R_{vir})/\msun \approx 3.2$, which is comparable to the silicon ion mass found in dwarf CGM by \citet{Zheng.etal.2024} for a nearby dwarf galaxy sample with a median stellar mass of $\langle\log M_\mathrm{star}/\msun\rangle_{\rm med} \approx 8$ but a factor of three greater than what is reported in the inner halo ($d_\mathrm{proj}/R_\mathrm{vir} \leq 0.15$) of the Local Group dwarf Sextans B by \citet{Fox:2026}. While the three ionization stages combined are commonly assumed to recover nearly the total underlying silicon abundance, up to 50\% of silicon may reside in higher ionization stages in photoionized gas with densities below $n_{\rm H}=0.001\,\cmjjj$ \citep[e.g.,][]{Chen:2017M}. Applying this ionization correction to the observed mass in \ion{Si}{2}–{\small IV}, we infer a total silicon mass of $\log M_\mathrm{Si}(<0.3,R_{\rm vir})/\msun \approx 3.5$ around a dwarf galaxy of $\log M_\mathrm{star}/\msun \approx 7.6$. Assuming a solar-like abundance pattern \citep[][]{Asplund.etal.2009}, this total silicon mass corresponds to a total metal mass of $\log M_Z(<0.3,R_{\rm vir})/\msun \approx 4.8$ in these low-mass halos.

The high-mass subsample with $\langle\log M_\mathrm{star}/\msun\rangle_{\rm med} = 8.6$ and $R_\mathrm{vir} = 78.9$ kpc has two of four sightlines with $N_\mathrm{Si} > 10^{14.5}\,\mathrm{cm}^{-2}$, corresponding to a covering fraction of $\kappa_{\rm Si} \sim 0.5$ for these high Si column density absorbers. We estimate a total enclosed mass of $\log M_\mathrm{Si\,II-IV}(<0.3\,R_{\rm vir})/\msun \approx 4.9$ or $M_\mathrm{Si}(<0.3\,R_{\rm vir})/\msun \approx 5.2$ after accounting for silicons in still higher ionization stages.  This is a factor of $\sim3-10$ higher than what was found in \citet[][]{Johnson.etal.2017} but similar to the Si mass inferred for the Local Group dwarf IC 1613 with$\mstar\approx 10^8\,M_\odot$ \citep[see e.g.,][]{Zheng.etal.2020}. This corresponds to $\log M_{Z}(<0.3\,R_{\rm vir})/\msun \approx 6.5$ for high-mass dwarf galaxies. 

\begin{figure*}
    \centering
    \includegraphics[width = \textwidth]{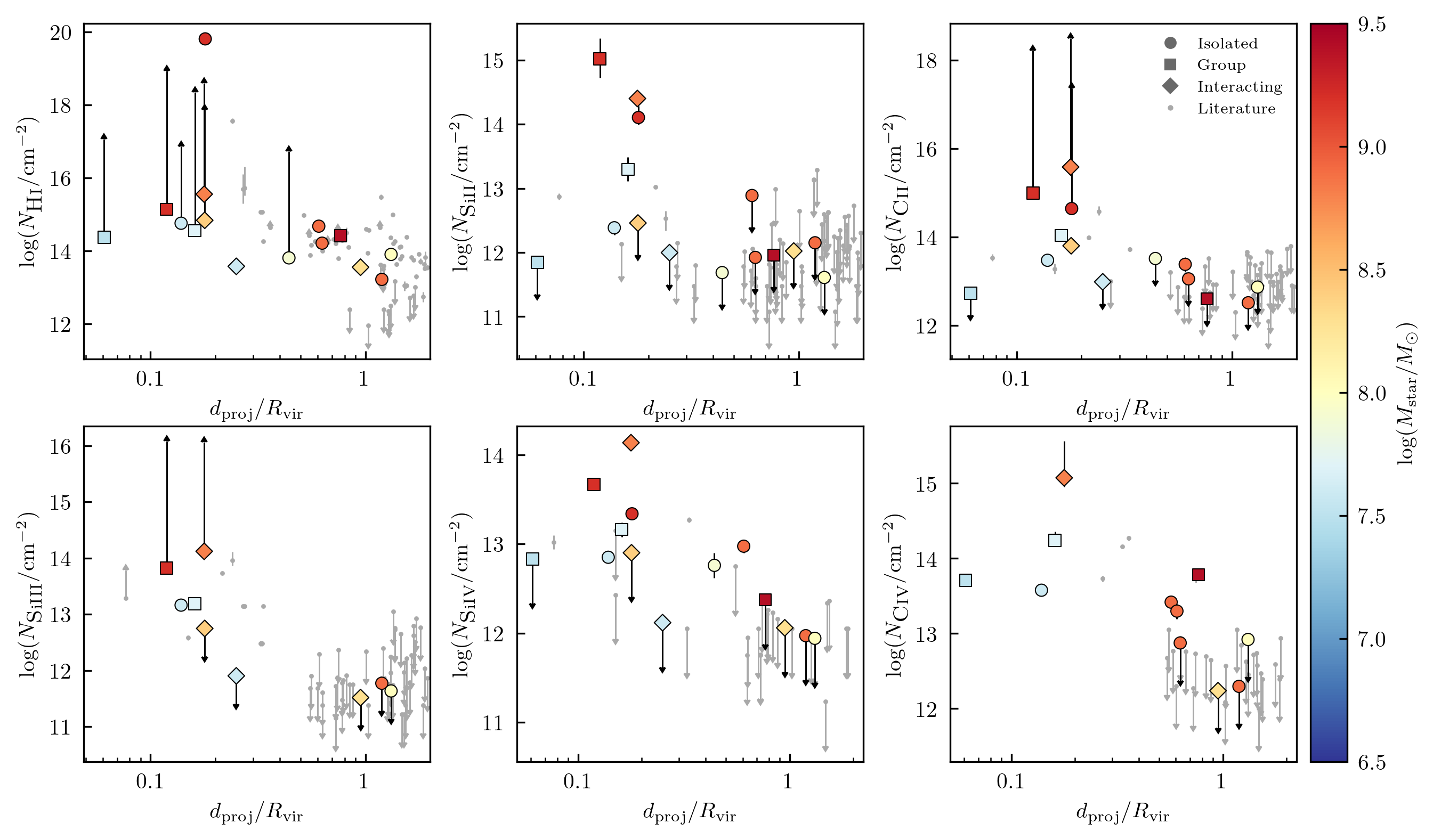}
    \caption{The column density of different ions along the sightlines near our first release subsample of QUEST Dwarf galaxies. With this preview of just $\sim20\%$ of our full sample, we are beginning to meaningfully populate the absorption properties of the inner CGM ($d_\mathrm{proj}/R_\mathrm{vir} \lesssim 0.5$) of dwarf galaxies. Non-Local Group literature data are from the high SNR ($S/N > 8$) subset of the \citet[][]{Zheng.etal.2024} compilation \citep{Bordoloi.etal.2014, Bordoloi.etal.2018, Liang.Chen.2014, Johnson.etal.2017}, \citet{Mishra.etal.2024}, and \citet{Johnson.etal.2026}.
    \label{fig:cgm_mass}}
\end{figure*}

Adopting the nucleosynthetic yields summarized in \citet[][]{Vincenzo.etal.2016}, we find that for low-mass dwarf galaxies, the cool CGM probed by low-ions contains 
$\approx 3$\% of the total metals produced over the lifetime of the galaxy; for high-mass dwarf galaxies, these numbers are a factor of $\sim5$ higher, with the inner cool CGM retaining $\approx 16$ \%
of the total  
metals produced by the galaxies. 
{We note, however, that the multiphase nature of the CGM \citep[e.g.,][]{Chen.Zahedy.2026} precludes a complete baryon census from any single ion. A global ionization gradient, in which low-ionization absorption declines more rapidly with radius than higher-ionization species \citep[e.g.,][see also Figure \ref{fig:cgm_mass} below]{Liang.Chen.2014}, requires multiple ionic tracers to probe the increasingly larger volume at greater halo radii. Indeed, \ion{O}{6} surveys have uncovered substantial reservoirs in the outer CGM and beyond, with the mass in O$^{5+}$ alone comparable to that traced by the low ions \citep[see][for a recent review]{McQuinn.etal.2026}. Given an O$^{5+}$ fraction of at most $\approx20$\%, these observations imply a substantially larger metal reservoir in the highly ionized CGM. At large radii, however, absorption measurements are increasingly susceptible to contamination from correlated halos projected along the line of sight \citep{Ho:2021}. Together, these considerations highlight the challenge of achieving a complete metal census across the full spatial and ionization extent of the CGM.}

We also examine the spatial distributions of individual ionic species. Figure~\ref{fig:cgm_mass} presents the column densities of \ion{H}{1}, \ion{Si}{2}, \ion{C}{2}, \ion{Si}{3}, \ion{Si}{4}, and \ion{C}{4} as a function of halo-radius-normalized projected distance, $d_{\rm proj}/R_\mathrm{vir}$, from top left to bottom right. The data points are color-coded by stellar mass, while analogous figures color-coded by $\log(\mathrm{H}\alpha/\mathrm{FUV})$ and gas-phase metallicity are presented in Appendix~\ref{app:ions}.

Consistent with previous studies of more massive halos, neutral hydrogen is detected throughout the halo and beyond $R_\mathrm{vir}$, indicating the presence of an extended reservoir of circumgalactic gas. In contrast, low-ionization species such as \ion{Si}{2}, \ion{C}{2}, and \ion{Si}{3} are preferentially found in the inner CGM at $d_{\rm proj}\lesssim0.3\,R_\mathrm{vir}$, where the largest column densities are observed \citep[e.g.,][]{Liang.Chen.2014}. Higher-ionization species, particularly \ion{Si}{4} and \ion{C}{4}, exhibit a substantially shallower radial decline and remain detectable to larger projected distances. Together, these trends point to a systematic ionization gradient within the CGM, in which cool, low-ionization gas is concentrated toward the inner halo while more highly ionized gas occupies a larger fraction of the halo volume \citep[see also][]{Chen:2001, Kong:2026}. Notably, the stellar-mass dependence is visible in nearly every ion, especially for \ion{Si}{2}, \ion{C}{2}, and \ion{C}{4}. 

In summary, our analysis suggests that both halo-scale and galaxy-scale processes govern the metal-bearing CGM of dwarf galaxies. The spatial distributions of individual ions reveal a multiphase structure, with low-ionization species concentrated toward the inner halo and higher-ionization species extending to larger projected distances. Galaxies with larger stellar masses also consistently exhibit stronger metal absorption at fixed $d_{\rm proj}/R_\mathrm{vir}$, indicating that stellar mass plays an important role in setting the overall metal content of the CGM. While there are tentative indications that recent star formation activity and gas-phase metallicity may also influence absorber strength, the current sample does not permit disentangling these effects from the strong covariance among galaxy mass, star formation, and chemical enrichment. The inferred metal masses should therefore be regarded as uncertain lower limits on the total CGM metal budget, particularly given its multiphase nature \citep[see, e.g.,][]{Faucher-Giguere.Oh.2023}. While \ion{Si}{2}–\ion{Si}{4} primarily trace cool, photoionized gas at $T\sim10^4$ K, more highly ionized species such as \ion{O}{6} probe lower-density and warmer gas \citep[e.g.,][]{Chen.Zahedy.2026} and may account for  $\approx$40\% of the metal budget around dwarf galaxies with $\log\,\mstar/\msun\approx8.4$ \citep[e.g.,][]{Johnson.etal.2017} with an approximately equal mass traced by \ion{O}{6} beyond the projected virial radius \citep[][]{Tchernyshyov:2022, Mishra.etal.2024}. 

At the same time, a continuous distribution of gas density and temperature likely produces some overlap between the material traced by low and high ions, complicating a simple accounting of metals across different phases. The full QUEST Dwarfs sample will substantially improve both the statistical power and parameter-space coverage needed to establish a more complete census of metals across the multiphase CGM, disentangle the roles of galaxy mass, star formation, and chemical enrichment, and distinguish among the physical processes that regulate the baryon cycle in dwarf galaxies.

\section{Conclusions} \label{sec:conclusions}

We present initial results on the CGM properties in low-mass halos using the first-release QUEST Dwarfs sample, which contains 14 out of the main sample of $>60$ dwarf galaxies with stellar mass $\mstar\le 10^9\,\msun$ at $z\approx 0.001$--0.16.  Every dwarf galaxy in the QUEST Dwarfs sample has at least one absorption probe of its circumgalactic medium at projected distances $d_{\rm proj}<100$ kpc. In addition, extensive multi-wavelength imaging and spectroscopic data are available for these galaxies from both public archives and our own observations to facilitate a comprehensive stellar population synthesis study for constraining the star formation history, stellar age, metallicity, and ISM gas phase metallicity. This initial ``first release'' sample is selected to demonstrate the extent of science possible with this integrated dataset.  We summarize the main findings below.

\begin{itemize}

    \item The ``first release'' sample is broadly representative of low-metallicity field dwarf galaxies. Eleven galaxies in the sample lie on the star-forming main sequence, while three are classified as quiescent based on their low specific star formation rates. The inclusion of the first-release QUEST Dwarfs increases by nearly a factor of three the number of available probes of the inner CGM ($d_{\rm proj}<0.3\,R_\mathrm{vir}$) in low-mass halos beyond the Local Group.
    
    \item The total silicon column density, combining \ion{Si}{2}, \ion{Si}{3}, and \ion{Si}{4}, declines with increasing halo-radius-normalized projected distance, indicating that chemically enriched gas is concentrated toward the inner CGM. In addition, at a fixed $d_{\rm proj}/R_\mathrm{vir}$, more massive galaxies tend to exhibit higher silicon column densities.
    
    \item The stellar-mass dependence is visible in nearly every ion, and is especially apparent for \ion{Si}{2}, \ion{C}{2}, and \ion{C}{4}. It suggests that both projected distance and galaxy mass play important roles in determining the metal content of the CGM, with distance governing the overall radial decline and stellar mass setting the normalization of the metal column density profile.
    
    \item The spatial distributions of individual ions reveal a multiphase structure, with low-ionization species concentrated toward the inner halo and higher-ionization species extending to larger projected distances. 
    
    \item While there are tentative indications that recent star formation activity and gas-phase metallicity may also influence absorber strength, the current sample does not permit disentangling these effects from the strong covariance among galaxy mass, star formation, and chemical enrichment. We anticipate that the full QUEST Dwarfs sample will substantially improve the statistical power and parameter-space coverage needed to disentangle these effects and distinguish among the different physical scenarios.

    \item Based on a preliminary division of literature and QUEST Dwarfs galaxies with \ion{Si}{2}–\ion{Si}{4} measurements into ``low-mass'' ($\log \mstar/M_\odot < 8$) and ``high-mass’’ ($\log \mstar/M_\odot \geq 8$) populations, we find that the cool CGM retains $\sim3$\% of the metals produced by low-mass dwarfs, compared to $\sim16$\% for high-mass dwarfs.
    
\end{itemize}

\begin{acknowledgements} 
AP thanks Andrey Kravtsov for sharing code to approximate the \citet[][]{Manwadkar.Kravtsov.2022} SHMR and Adam Carnall for his input on best using \texttt{Bagpipes}. The authors thank Sean Johnson for comments and discussion on this manuscript.
AP was supported in part by the Quad Fellowship administered by IIE.  HWC and AP acknowledge partial support from HST-AR-17049.007.A, 
provided through a grant from the STScI under NASA contract NAS5-26555.
EB acknowledges support from NASA under award number 80GSFC24M0006.
TKC was supported by the E. Margaret Burbidge Prize Postdoctoral Fellowship from the Brinson Foundation at the Departments of Astronomy and Astrophysics at the University of Chicago.
This research is based on observations made with the NASA/ESA \textit{Hubble Space Telescope} obtained from the Space Telescope Science Institute, which is operated by the Association of Universities for Research in Astronomy, Inc., under NASA contract NAS 5–26555. These observations are associated with programs 9124 (PI: Windhorst), 11520 (PI: Green), 11585 (PI: Crighton), 11598 (PI: Tumlinson), 12025 (PI: Green), 12038 (PI: Green), 12486 (PI: Bowen), 12546 (PI: Tully), 12603 (PI: Heckman), 13008 (PI: Stocke), 13024 (PI: Mulchaey), 13314 (PI: Borthakur), 13852 (PI: Bordoloi), 14071 (PI: Borthakur), 14772 (PI: Wakker), 14777 (PI: Arav), and 17070 (PI: Kilpatrick). \textit{GALEX} imaging comes from both the public surveys and guest investigator programs 25 (PI: Kennicutt), 40 (PI: Barstow), and 104 (PI: Donovan).
This paper includes data gathered with the 6.5 meter Magellan Telescopes located at Las Campanas Observatory, Chile. MMT Observatory access was supported by Northwestern University and the Center for Interdisciplinary Exploration and Research in Astrophysics (CIERA).
This research has also made use of the Keck Observatory Archive (KOA), which is operated by the W. M. Keck Observatory and the NASA Exoplanet Science Institute (NExScI), under contract with the National Aeronautics and Space Administration. These data are from Program IDs U050LA (PI: Prochaska) and U21L (PI: Bolte).
\end{acknowledgements}

\appendix 

\renewcommand{\thefigure}{A\arabic{figure}}
\setcounter{figure}{0}
\renewcommand{\thetable}{A\arabic{table}}
\setcounter{table}{0}

\section{Photometric and spectral properties of QUEST Dwarfs} \label{app:photspec}

\begin{figure*}
    \centering
    \includegraphics[width = 0.75\linewidth]{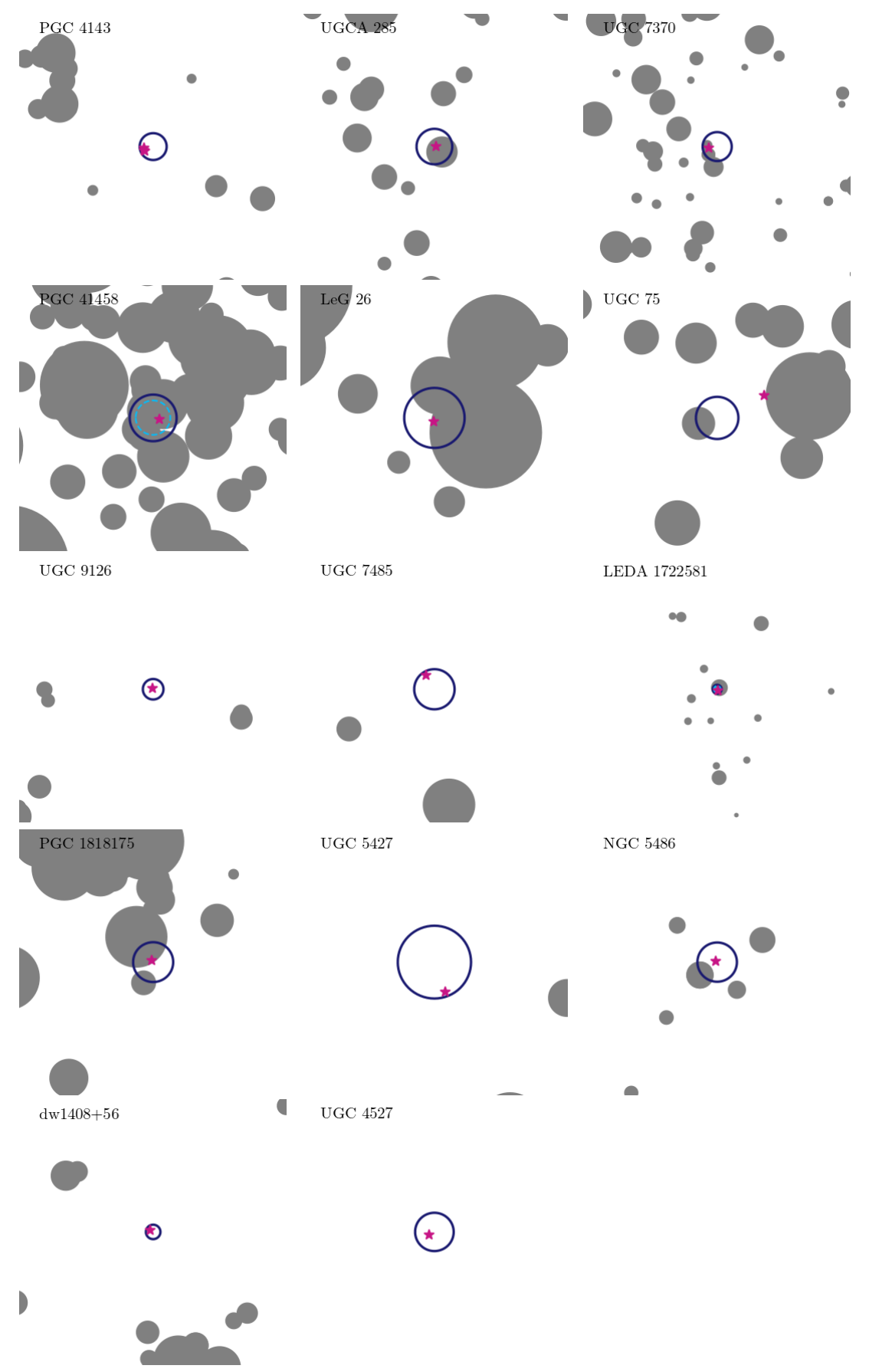}
    \caption{The cosmic ``neighborhood'' of each QUEST Dwarf. Each panel spans $5\times5$ deg$^2$ and shows in gray the virial coverage of all galaxies within $\pm 300$ km s$^{1}$ as selected from the 50 MGC  \citep{Ohlson.etal.2024} and DESI DR1 \citep{DESI.DR1} catalogs. The virial radius of the each QUEST Dwarf is shown in navy blue and QSO sightlines are denoted with pink stars. For PGC 41458 and LEDA 1722581, we mark the virial radius of their interacting companions with light blue dashed circles.
    \label{fig:env}}
\end{figure*}

As discussed in \S\,\ref{sec:galanalysis}, we measure broadband FUV and optical photometry of the sample galaxies based on a 2D surface brightness profile analysis of archival imaging data.  We also measure nebular line fluxes and equivalent widths, along with the strength of the 4000~\AA\ discontinuity, using available optical spectra.  All the measurements are presented in Tables \ref{tab:phot} and \ref{tab:spec} below.  The environmental properties of each galaxy (see \S\,\ref{subsec:environ}) are presented in Figure \ref{fig:env}.

\begin{rotatetable*}{l|ccccccccc}
    \tablecaption{Photometric properties of the galaxies in the QUEST Dwarfs first release sample as obtained from survey imaging. All magnitudes AB. \label{tab:phot}}
    \tablewidth{\linewidth}
    \tablehead{\textbf{}  &  \textbf{\textit{g}} & \textbf{\textit{r}} & \textbf{\textit{z}} & \textbf{FUV} & 
    \textbf{n} & \textbf{b/a} & \textbf{PA (deg)} & \textbf{\boldmath r$_\mathrm{eff}$\unboldmath~(arcsec)} \\
    \textbf{Galaxy}  &  (1) & (2) & (3) & (4) & 
    (5) & (6) & (7) & (8)}
    \startdata
    \textbf{PGC 4143} & $14.682\pm0.002$ & $14.140\pm0.003$ & $13.784\pm0.003$ & $17.72\pm0.02$ & $1.660\pm0.005$ & $0.5802{+0.0009\atop-0.0008}$ & $-14.46{+0.04\atop-0.05}$ & $26.91\pm0.07$\\
    \textbf{UGCA 285} & $16.261{+0.003\atop-0.002}$ & $15.85{+0.03\atop-0.05}$ & $15.61\pm0.04$ & $19.4\pm0.1$ & $0.655\pm0.008$ & $0.881{+0.008\atop-0.011}$ & $8{+1\atop-2}$ & $19.2{+0.6\atop-0.8}$\\
    \textbf{UGC 7370} & $14.486\pm0.001$ & $13.940\pm0.002$ & $13.576\pm0.002$ & $17.49{+0.06\atop-0.05}$ & $0.817\pm0.002$ & $0.2828\pm0.0003$ & $-29.37\pm0.02$ & $20.31{+0.06\atop-0.05}$\\
    \textbf{PGC 41458} & $16.725\pm0.002$ & $16.095\pm0.004$ & $15.719\pm0.004$ & $22.1\pm0.1$ & $1.354\pm0.006$ & $0.574\pm0.002$ & $-27.4{+0.1\atop-0.2}$ & $13.29\pm0.03$\\
    \textit{PGC 41458b} & $17.78\pm0.01$ & $17.19{+0.02\atop-0.04}$ & $16.67\pm0.01$ & $>23.8$ & $0.71\pm0.01$ & $0.97{+0.01\atop-0.04}$ & $-24{+4\atop-6}$ & $12\pm1$\\
    \textbf{LeG 26} & $16.403\pm0.002$ & $15.798{+0.006\atop-0.008}$ & $15.419\pm0.007$ & $21.483\pm0.003$ & $0.884\pm0.003$ & $0.811{+0.004\atop-0.005}$ & $79.7\pm0.5$ & $14.28{+0.06\atop-0.07}$\\
    \textbf{UGC 75}\tablenotemark{a} & $12.8458{+0.0006\atop-0.0005}$ & $12.359\pm0.003$ & $12.015{+0.005\atop-0.006}$ & ... & $1.251\pm0.004$ & $0.547\pm0.001$ & $26.02\pm0.06$ & $26.7\pm0.3$ \\
    &  &  &  &  & $0.652\pm0.004$ & $0.9994{+0.0003\atop-0.0006}$ & $26.02\pm0.06$ & $42\pm4$ \\
    \textbf{UGC 9126} & $14.924\pm0.001$ & $14.526\pm0.003$ & $14.279\pm0.004$ & $16.92\pm0.05$ & $1.000\pm0.003$ & $0.719\pm0.002$ & $1.5\pm0.2$ & $30.38{+0.09\atop-0.10}$ \\
    \textbf{UGC 7485} & $14.907\pm0.002$ & $14.387$ & $14.078\pm0.003$ & $17.90{+0.08\atop-0.07}$ & $1.478\pm{+0.005\atop-0.004}$ & $0.671\pm0.002$ & $-59.69{+0.08\atop-0.09}$ & $14.46{+0.05\atop-0.04}$ \\
    \textbf{LEDA 1722581} & $17.549\pm0.003$ & $17.085\pm0.004$ & $16.800\pm0.005$ & $21.4{+0.3\atop-0.2}$ & $1.157\pm0.008$ & $0.326\pm0.001$ & $-82.16\pm0.08$ & $6.56\pm0.02$\\
    \textit{LEDA 1722581-em}\tablenotemark{b} & ... & ... & ... & $20.6\pm0.2$ & ... & ... & ... & ...\\ 
    \textbf{PGC 1818175} & $15.734\pm0.002$ & $15.137\pm0.003$ & $14.751\pm0.003$ & $21.6{+0.5\atop-0.3}$ & $1.145\pm0.003$ & $0.490\pm0.001$ & $-76.49\pm0.06$ & $17.73\pm0.03$\\
    \textbf{UGC 5427} & $14.669\pm0.002$ & $14.034\pm0.004$ & $14.084\pm0.004$ & $16.517\pm0.009$ & $0.723{+0.001\atop-0.002}$ & $0.611\pm0.002$ & $-76.49 \pm 0.06$ & $19.20\pm0.06$ \\
    \textbf{NGC 5486}\tablenotemark{a} & $13.7678{+0.0010\atop-0.0007}$ & $13.326{+0.002\atop-0.003}$ & $13.328{+0.005\atop-0.004}$ & $15.598\pm0.006$ & $0.65\pm0.01$ & $0.573\pm0.002$ & $70.5\pm0.2$ & $25.2\pm0.4$ \\
     &  &  &  &  & $3.81{+0.04\atop-0.05}$ & $0.99995{+0.00004\atop-0.00009}$ & $70.5\pm0.2$ & $10\pm1$ \\
    \textbf{dw1408$+$56} & $17.977\pm0.005$ & $17.551{+0.009\atop-0.014}$ & $17.96\pm0.02$ & $19.7{+0.2\atop-0.1}$ & $0.925\pm0.008$ & $0.569{+0.005\atop-0.006}$ & $51.6\pm0.4$ & $12.52\pm0.06$ \\
    \textbf{UGC 4527} & $16.158{+0.006\atop-0.007}$ & $15.755{+0.008\atop-0.009}$ & $16.150\pm0.009$ & $18.35{+0.09\atop-0.08}$ & $0.805\pm0.003$ & $0.814\pm0.002$ & $30.0\pm0.3$ & $14.71\pm0.08$
    \enddata
    \tablenotetext{a}{These galaxies are best fit by the combination of two S\'{e}rsic models with a fixed PA. We list parameters for both models here, with $g$, $r$, $z$, and FUV magnitudes coming from the combined profile. The top row includes the parameters for the primary S\'{e}rsic profile, which, for UGC 75 and NGC 5486 contain $(77.8{+0.7\atop-0.8})\%$ and $(71.5\pm 0.6)\%$ of the flux respectively, and the bottom row includes parameters for the secondary profile.}
    \tablenotetext{b}{The FUV magnitude is determined from the same region used in extracting the MaNGA spectrum. LEDA 1722581 is correspondingly masked, so FUV flux is attributed to only one of LEDA 1722581 or LEDA 1722581-em.}
\end{rotatetable*}

\begin{rotatetable*}{l|cccccccc}
    \tablecaption{Individual spectral line and index measurements for the QUEST Dwarfs first release sample. For galaxies spectra from with multiple independent pointings, reported values are summed. \label{tab:spec}}
    \tablewidth{\linewidth}
    \tablehead{ \textbf{Galaxy} &  \textbf{\boldmath[\textsc{Oii}]$_{3728}$ (erg s$^{-1}$)\unboldmath} & \textbf{\boldmath$\mathrm{H\beta}_{4862}$ (erg s$^{-1}$)\unboldmath} &  \textbf{\boldmath[\textsc{Oiii}]$_{4960}$ (erg s$^{-1}$)\unboldmath} & \textbf{\boldmath[\textsc{Oiii}]$_{5008}$ (erg s$^{-1}$)\unboldmath} & \textbf{\boldmath$\mathrm{H\alpha}_{6564}$ (erg s$^{-1}$)\unboldmath} & \textbf{\boldmath[\textsc{Nii}]$_{6585}$ (erg s$^{-1}$)\unboldmath} & \textbf{\boldmath$D_{n}4000$\unboldmath}\\
    {}  &  \textbf{\boldmath${\rm EW}_{3728}$ (\AA)\unboldmath} & \textbf{\boldmath${\rm EW}_{4862}$ (\AA)\unboldmath} &  \textbf{\boldmath${\rm EW}_{4960}$ (\AA)\unboldmath} & \textbf{\boldmath${\rm EW}_{5008}$ (\AA)\unboldmath} & \textbf{\boldmath${\rm EW({H\alpha}})$ (\AA)\unboldmath} & \textbf{\boldmath${\rm EW}_{6585}$ (\AA)\unboldmath} & {}}
    \startdata
    \label{tab:lines}
    \textbf{PGC 4143} & $(1.62 \pm 0.08)\times 10^{-14}$ & ... & $(2.2\pm0.1)\times10^{-15}$ & $(5.2\pm0.1)\times10^{-15}$ & $(1.53\pm0.01)\times10^{-14}$ & $(2.5\pm0.1)\times10^{-15}$ & $1.17\pm0.01$\\
    & $14.0\pm0.8$ & ... & $1.20\pm0.08$ & $2.92\pm0.08$ & $9.25\pm0.08$ & $1.51\pm0.07$ & \\
    \textbf{UGCA 285\tablenotemark{a}} & $\mathit{(1.5\pm0.3)\times10^{-14}}$ & $\mathit{(0.3\pm1.0)\times10^{-15}}$ & n.d & $\mathit{(0.9\pm1.8)\times10^{-15}}$ & $(1.34\pm0.06)\times10^{-14}$ & $(4.1\pm0.4)\times10^{-15}$ & $\mathit{0.74\pm0.05}$\\
    & $27\pm7$ & $0.5\pm1.7$ & n.d. & $2\pm3$ & $23\pm2$ & $6.9\pm0.3$ & \\
    \textbf{UGC 7370} & $(1.59\pm0.03)\times10^{-13}$ & ... & $(3.30\pm0.07)\times10^{-14}$ & $(6.96\pm0.06)\times10^{-14}$ & $(2.198\pm0.004)\times10^{-13}$ & $(3.78\pm0.04)\times10^{-14}$ & $1.080 \pm 0.008$\\
    & $20.1\pm0.5$ & ... & $3.16\pm0.07$ & $6.68\pm0.07$ & $23.5\pm0.2$ & $4.05\pm0.05$ & \\
    \textbf{PGC 41458} & $(1.3\pm0.5)\times10^{-15}$ & $(5\pm2)\times10^{-16}$ & $(3\pm1)\times10^{-16}$ & $(2\pm1)\times10^{-16}$ & $(3\pm2)\times10^{-16}$ & $(8\pm5)\times10^{-17}$ & $1.04 \pm 0.04$\\
    & $1.1\pm0.4$ & $0.5\pm0.2$ & $0.3\pm0.1$ & $0.2\pm0.1$ & $0.355\pm0.02$ & $0.09\pm0.06$ & \\
    \textit{PGC 41458b} & $(1.1\pm0.3)\times10^{-14}$ & $(-6\pm2)\times10^{-16}$ & $(4\pm1)\times10^{-16}$ & $(0.7\pm1.6)\times10^{-16}$ & $(8\pm7)\times10^{-17}$ & $(2.1\pm0.5)\times10^{-16}$ & $0.9\pm0.1$\\
    & $16\pm5$ & $-3\pm1$ & $2.3\pm0.7$ & $4\pm1$ & $0.5\pm1.1$ & $1.5\pm0.4$ & \\
    \textbf{LeG 26\tablenotemark{a}} & ... & $\mathit{(-1.0\pm0.5)\times10^{-16}}$ & $\mathit{(1.3\pm0.4)\times10^{-16}}$ & $\mathit{(3.6\pm0.8)\times10^{-16}}$ & $(7.8\pm0.6)\times10^{-16}$ & $(9\pm2)\times10^{-17}$ & ...\\
    & ... & $-0.3\pm0.2$ & $0.4\pm0.1$ & $1.2\pm0.3$ & $3.5\pm0.3$ & $0.4\pm0.1$ & \\
    \textbf{UGC 75} & $(6.3\pm0.1)\times10^{-14}$ & ... & $(1.37\pm0.02)\times10^{-14}$ & $(4.25\pm0.02)\times10^{-14}$ & $(9.94\pm0.02)\times10^{-14}$ & $(1.47\pm0.01)\times10^{-14}$ & $1.126\pm0.004$ \\
    & $14.5\pm0.3$ & ... & $2.04\pm0.03$ & $6.34\pm0.05$ & $18.56\pm0.08$ & $2.75\pm0.03$ &  \\
    \textbf{UGC 9126}\tablenotemark{b} & $(3.9\pm0.2)\times10^{-14}$ & $(1.04\pm0.04)\times10^{-14}$ & $(1.24\pm0.04)\times10^{-14}$ & $(3.89\pm0.04)\times10^{-14}$ & $(4.21\pm0.03)\times10^{-14}$ & $(1.9\pm0.2)\times10^{-15}$ & $0.96 \pm 0.04$ \\
    & $18\pm3$ & $9.3\pm0.6$ & $11.1\pm0.7$ & $34.4\pm0.8$ & $54\pm1$ & $2.5\pm0.6$ &  \\
    \textbf{UGC 7485} & ... & $(9.82\pm0.05)\times10^{-14}$ & $(1.145\pm0.005)\times10^{-13}$ & $(3.498\pm0.005)\times10^{-13}$ & $(3.756\pm0.005)\times10^{-13}$ & $(3.99\pm0.05)\times10^{-14}$ & ...\\
    & ... & $14.3\pm0.3$ & $16.4\pm0.3$ & $49.4\pm0.8$ & $67\pm1$ & $7.1\pm0.2$ &  \\
    \textbf{LEDA 1722581} & $(2.87\pm0.02)\times10^{-15}$ & $(6.72\pm0.05)\times10^{-16}$ & $(4.05\pm0.06)\times10^{-16}$ & $(1.307\pm0.006)\times10^{-15}$ & $(2.582\pm0.005)\times10^{-15}$ & $(3.74\pm0.04)\times10^{-16}$ & $1.195\pm0.003$\\
    & $23\pm0.5$ & $3.35\pm0.03$ & $2.02\pm0.03$ & $6.51\pm0.04$ & $15.5\pm0.1$ & $2.24\pm0.03$ & \\
    \textit{LEDA 1722581-em} & $(3.68\pm0.01)\times10^{-15}$ & $(1.032\pm0.004)\times10^{-15}$ & $(6.60\pm0.04)\times10^{-16}$ & $(2.009\pm0.004)\times10^{-15}$ & $(3.344\pm0.004)\times10^{-15}$ & $(3.83\pm0.03)\times10^{-16}$ & $1.102\pm0.003$\\
    & $49.2\pm0.9$ & $0.43\pm0.08$ & $6.70\pm0.05$ & $20.3\pm0.1$ & $42.4\pm0.3$ & $4.87\pm0.04$ & \\
    \textbf{PGC 1818175} & ... & $(2\pm2)\times10^{-17}$ & $(1.1\pm0.5)\times10^{-17}$ & $(4\pm1)\times10^{-17}$ & $(8\pm1)\times10^{-17}$ & $(4\pm2)\times10^{-17}$ & $1.08\pm0.03$\\
    & ... & $0.3\pm0.2$ & $0.15\pm0.07$ & $(0.6\pm0.2)$ & $1.4\pm0.2$ & $(0.6\pm0.3)$ & \\
    \textbf{UGC 5427} & $>1.33\times10^{15}$ & $(1.73\pm0.04)\times10^{-15}$ & $(4.0\pm0.3)\times10^{-16}$ & $(1.45\pm0.04)\times10^{-15}$ & $(4.65\pm0.06)\times10^{-15}$ & $(4.6\pm0.2)\times10^{-16}$ & $1.01\pm0.02$ \\
    & $24\pm2$ & $11.7\pm0.6$ & $2.7\pm0.6$ & $9.9\pm0.7$ & $46\pm1$ & $4.5\pm0.6$ &   \\
    \textbf{NGC 5486} & $>5.35\times10^{-15}$ & $(6.24\pm0.07)\times10^{-15}$ & $(6.44\pm0.07)\times10^{-15}$ & $(1.94\pm0.02)\times10^{-14}$ & $(1.92\pm0.01)\times10^{-14}$ & $(1.62\pm0.03)\times10^{-15}$ & $0.94\pm0.04$ \\
    & $51\pm2$ & $43.9\pm0.9$ & $46\pm2$ & $138\pm3$ & $245\pm4$ & $20.7\pm0.8$ &  \\
    \textbf{dw1408$+$56} & ... & $(8.3\pm0.7)\times10^{-16}$ & $(2.6\pm0.7)\times10^{-16}$ & $(9.0\pm0.6)\times10^{-16}$ & $(4.07\pm0.09)\times10^{-15}$ & $(7\pm1)\times10^{-16}$ & $1.18\pm0.02$ \\
    & ... & $5.6\pm0.4$ & $1.8\pm0.5$ & $6.2\pm0.4$ & $30\pm1$ & $5.6\pm0.9$ &  \\
    \textbf{UGC 4527} & ... & $(1.0\pm0.1)\times10^{15}$ & $(-6\pm1)\times10^{-16}$ & $(-3\pm1)\times10^{-16}$ & $(9.4\pm0.1)\times10^{-15}$ & $(9\pm1)\times10^{-16}$ & $1.29\pm0.01$\\
    & ... & $1.6\pm0.2$ & $-1.0\pm0.2$ & $-0.5\pm0.2$ & $19.2\pm0.4$ & $1.8\pm0.3$ &
    \enddata

    \tablecomments{Lines that are not included in the wavelength coverage of the spectra are denoted ``...''}
    \tablenotetext{a}{Measurements of line \textit{fluxes} at $\lambda < 5600$\AA~and the $D_n4000$ are significantly less robust for these galaxies due to the quality of the blue-side flux calibration from archival LRIS data. These numbers are italicized in the table.}
    \tablenotetext{b}{[\textsc{Nii}] is not detected in one of the two UGC 9126 pointings, so the value reported here is derived only from the spectrum along the major axis. N$_2$ is therefore computed only from the major axis spectrum for the metallicity estimate of this galaxy.}
\end{rotatetable*}

\section{Full spectral fitting} \label{app:spec}

\begin{figure*}
    \centering
    \includegraphics[width = \linewidth]{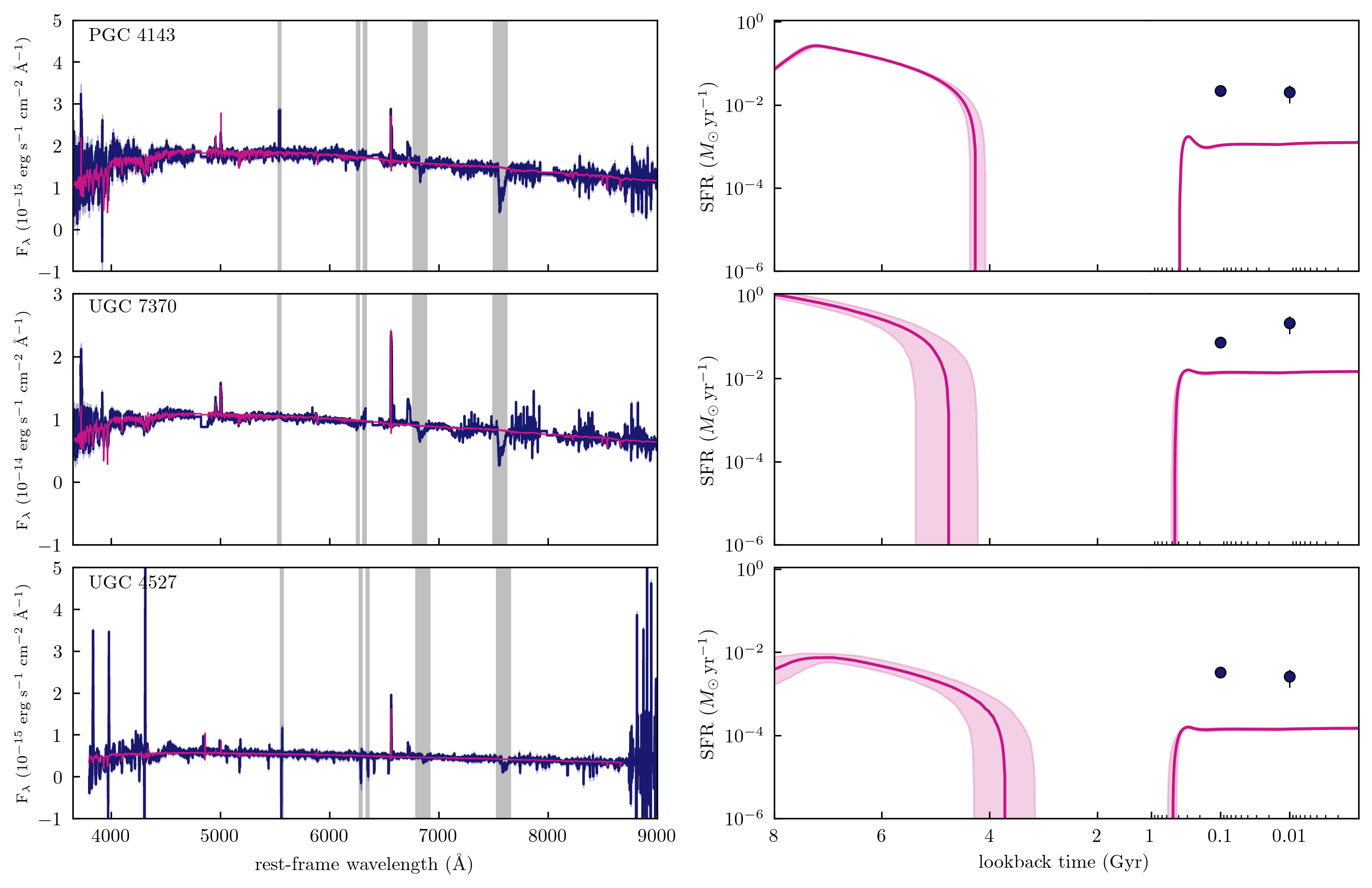}
    \caption{\emph{Left:} Optical spectra (blue) and \texttt{Bagpipes} best fit model spectra (pink) for galaxies with a ``bursty'' SFH. Note that the scaling is different spectrum-to-spectrum. Regions of the spectrum that are masked due to atmospheric features or sky lines are grayed out. \emph{Right:} Star formation histories recovered from \texttt{Bagpipes} fitting (pink); where the measurements are available, we also show the H$\alpha$ and FUV SFRs (see \S\  \ref{sec:sf}) on timescales of 10 and 100 Myr respectively.
    \label{fig:specsfhb}}
\end{figure*}

\begin{figure*}
    \centering
    \includegraphics[width = \linewidth]{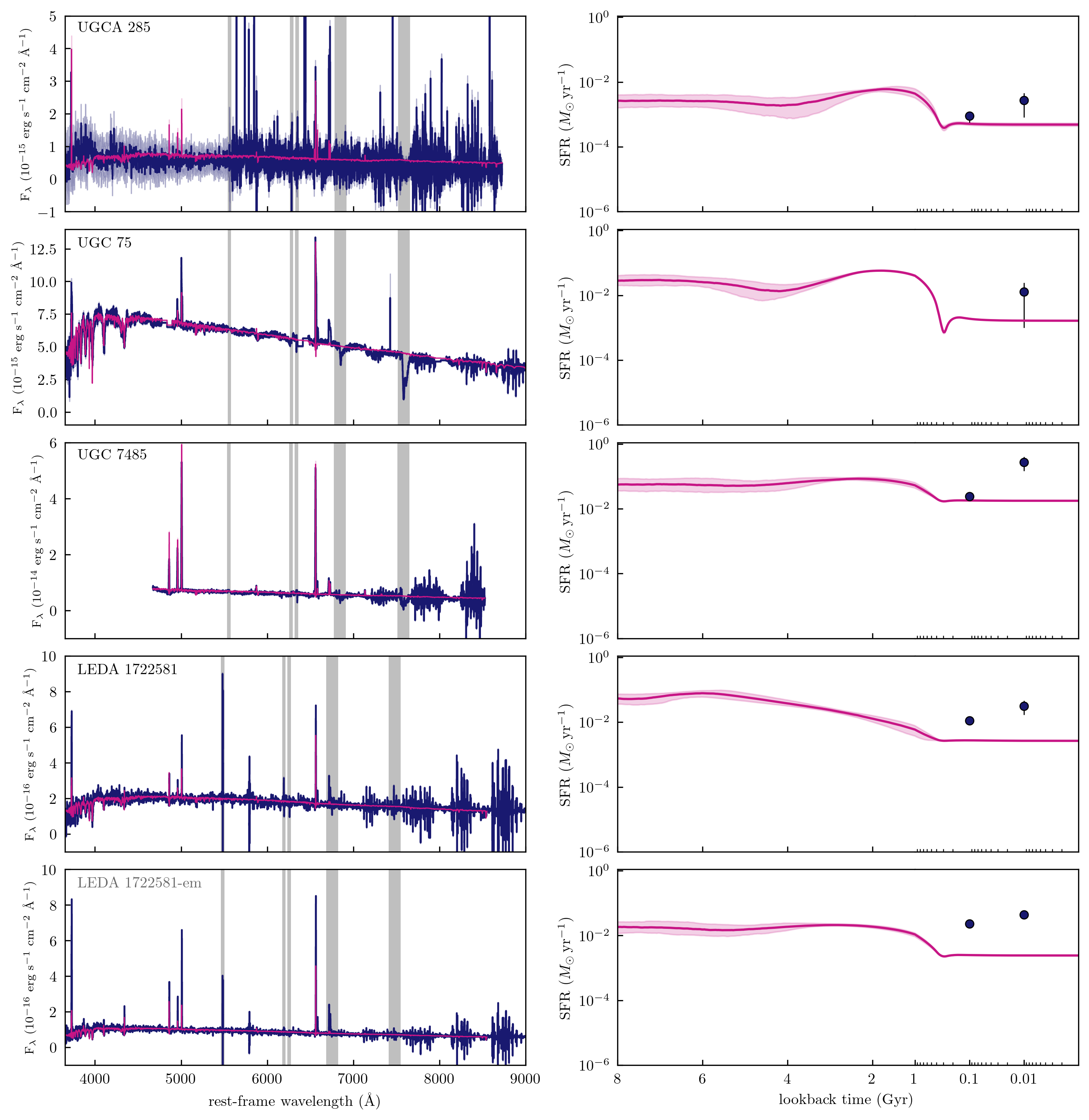}
    \caption{Same as Figure \ref{fig:specsfhb}, for galaxies with an ``sustained'' SFH. For galaxies with multiple spectroscopic pointings, we select a representative spectrum and fit to display here.
    \label{fig:specsfhs}}
\end{figure*}

\begin{figure*}
    \centering
    \includegraphics[width = \linewidth]{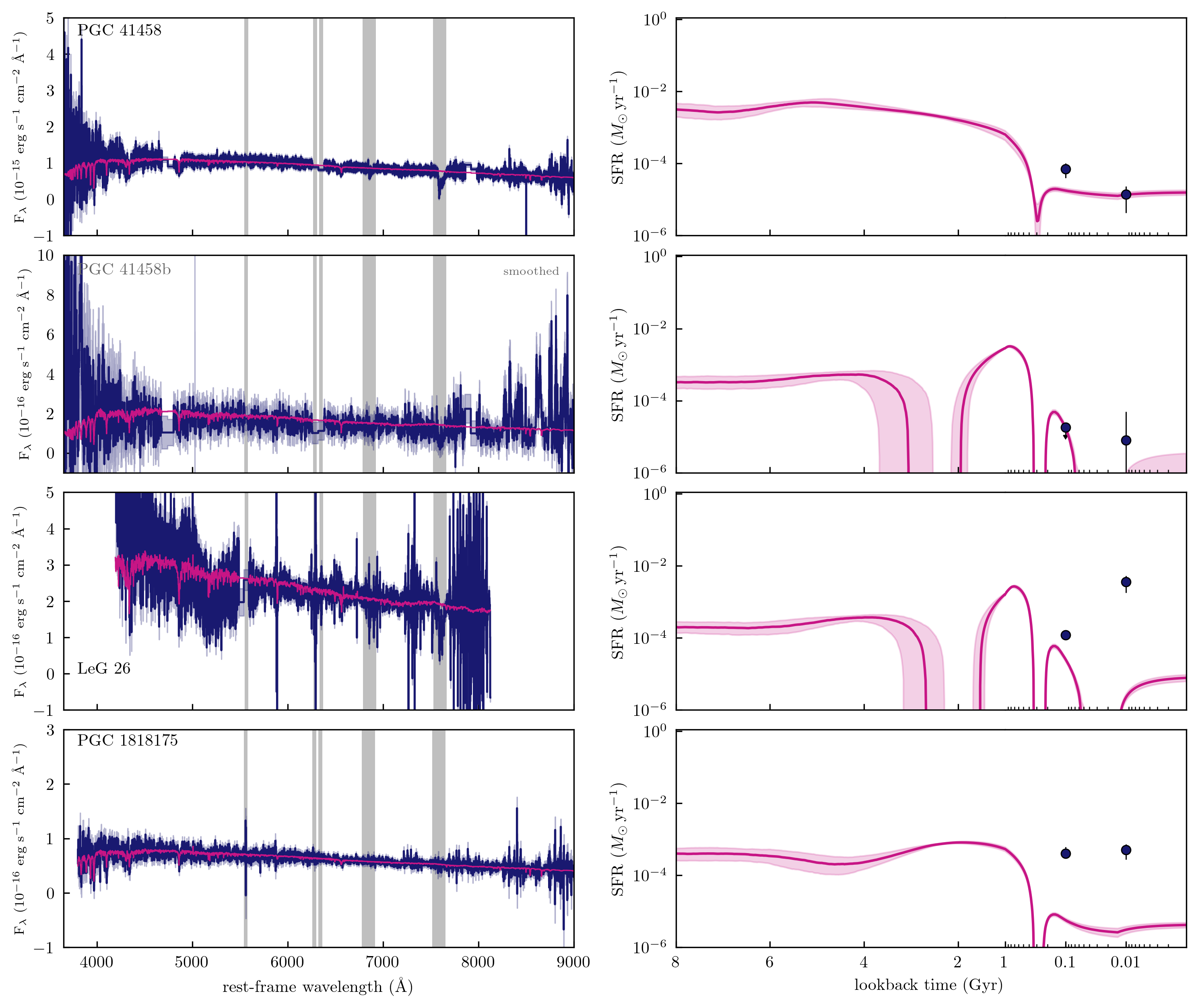}
    \caption{Same as Figure \ref{fig:specsfhb}, for galaxies with an ``quiescent'' SFH. For galaxies with multiple spectroscopic pointings, we select a representative spectrum and fit to display here.
    \label{fig:specsfhq}}
\end{figure*}

\begin{figure*}
    \centering
    \includegraphics[width = \linewidth]{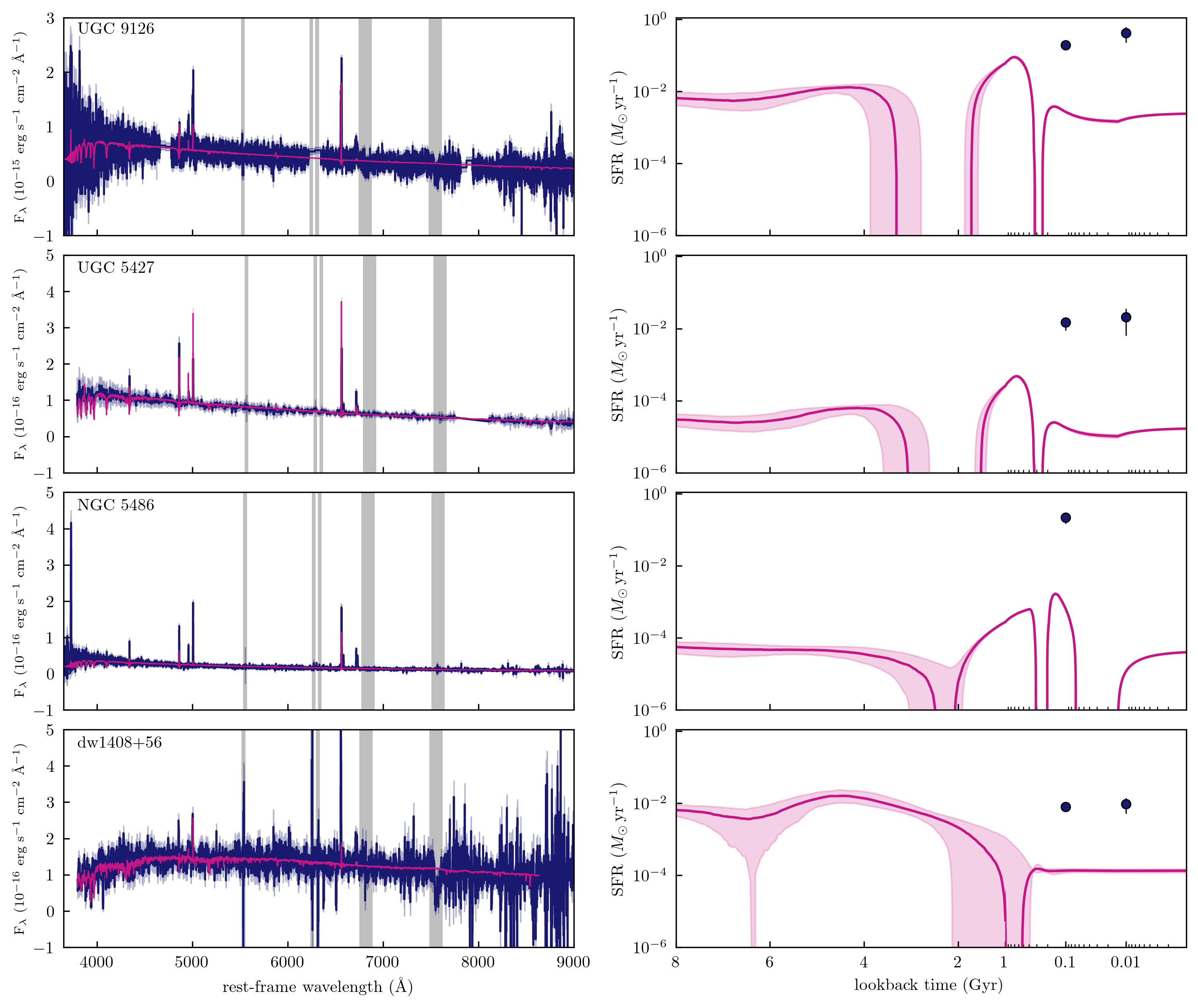}
    \caption{Same as Figure \ref{fig:specsfhb}, for galaxies with an ``episodic'' SFH. For galaxies with multiple spectroscopic pointings, we select a representative spectrum and fit to display here. Note that there is no correction made for slit or fiber loss, so the \texttt{Bagpipes} SFH for UGC 5427 and NGC 5486 deviates particularly from the ``integrated'' H$\alpha$ and FUV SFRs.
    \label{fig:specsfhe}}
\end{figure*}

We use \texttt{Bagpipes} to fit the full spectrum of each QUEST Dwarf galaxy, using a non-parametric star formation history model. The results of that fitting are shown in Figures \ref{fig:specsfhb} through \ref{fig:specsfhe}. The fit star formation histories are shown at right, where they offer \textit{qualitative} insight into the galaxies' evolution. For instance, LEDA 1722581 and its H$\alpha$-selected counterpart appear to have evolved in parallel despite the latter's lack of independent stellar continuum, and PGC 41458 and PGC 41458b have comparable star formation histories, pointing to their co-evolution, though the lower mass satellite PGC 41458b has a more stochastic SFH. Similarly, UGC 75 and UGC 7485, which are both shell galaxies, have consistent inferred star formation histories.

The choice of $\alpha = 5$ for the Dirichlet prior on the star formation history, which we adopt as optimal from \citet{Iyer.etal.2019}, effectively smooths the SFH, leading to much less bursty behavior among our seven time bins than may be suggested from FUV and H$\alpha$ SFR indicators. As such, \textit{quantitative} analysis of galaxies' SFHs comes from the direct photometric (FUV) and line (H$\alpha$ and $D_n4000$) measurements. We focus only show the last 8 Gyr of each galaxy's SFH, where a combination of absorption features (including the Balmer series) and emission lines best constrain stellar population ages and ongoing star formation.

\section{Photometric color conversions for SBF} \label{app:SBF}

Following e.g., \citet{Cohen.etal.2018} and \citet{Carlsten.etal.2019b}, we compare photometry for simple stellar populations in \textit{HST}/ACS, \textit{HST}/WFPC2, CFHT MegaCam, and DECam bands using photometry from MIST isochrones \citep{Dotter.2016, Choi.etal.2016}. We use the AB magnitude color for every MIST SSP with an age $\leq 13$ Gyr, spanning $-4 \leq$ [Fe/H] $\leq 0.5$, to directly relate DECam $g-z$ to each of \textit{HST}/ACS WFC F475W-F814W, \textit{HST}/WFPC2 F555W-F814W, and CFHT MegaCam $g-i$ to predict $\bar{M}$ for comparison with the magnitude of the SBF signal \citep{Ajhar.etal.1997, Blakeslee.etal.2010, Carlsten.etal.2019b}.

\begin{figure}
    \centering
    \includegraphics[width = \linewidth]{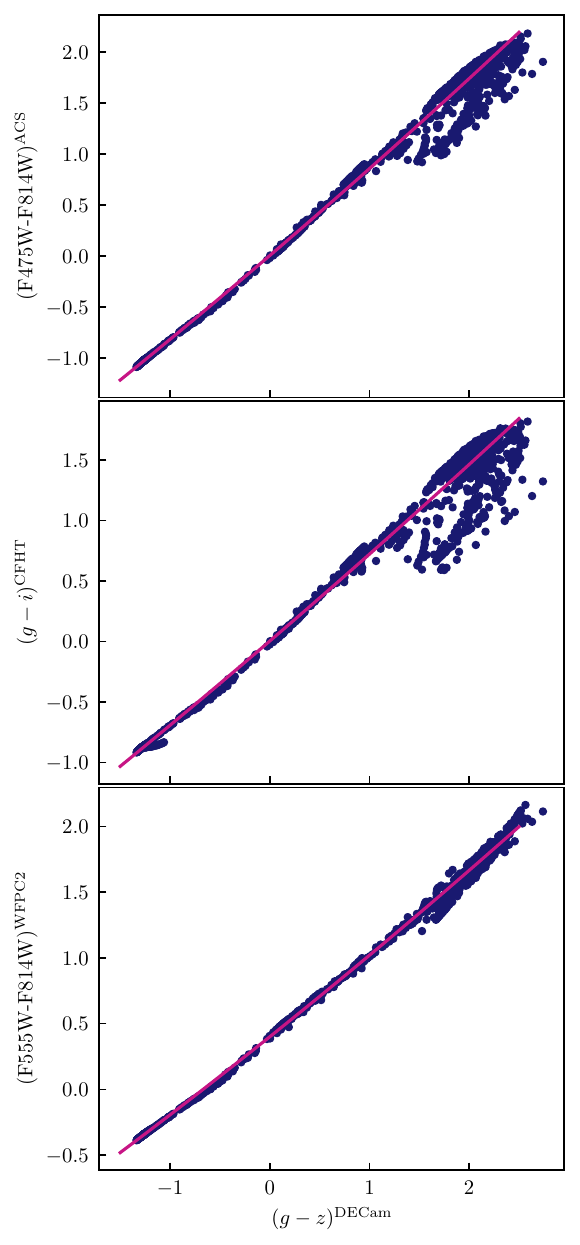}
    \caption{Synthetic photometry from MIST for simple stellar populations at a variety of ages ($10^5$ yr - 13 Gyr) and metallicities ($-4 \leq$ [Fe/H] $\leq 0.5$). The color indices of each SSP are shown as blue dots, and our fit to these points is shown in pink (Eqs. \ref{eq:hst-decals}, \ref{eq:cfht-decals}, and \ref{eq:wfpc2-decals}). \emph{Top:} \textit{HST}/ACS F475W-F814W compared to DECam $g-z$. \emph{Middle:} CFHT MegaCam $g-i$ compared to DECam $g-z$. \emph{Bottom:} \textit{HST}/WFPC2 F555W-F814W compared to DECam $g-z$.
    \label{fig:color}}
\end{figure}

As dwarf galaxies are generally bluer, we explicitly fit for all SSPs with $(g-z)^\mathrm{DECam} \leq 1.5$ (Figure \ref{fig:color}). We find the following color conversions:
\begin{equation}
    (\mathrm{F475W-F814W})^\mathrm{ACS} = 0.0154x^2 + 0.8366x + 0.0034 \label{eq:hst-decals}
\end{equation}
and 
\begin{equation}
    (g-i)^\mathrm{CFHT} = 0.0113x^2 + 0.7078x + 0.0033 \label{eq:cfht-decals}
\end{equation}
and 
\begin{equation}
    (\mathrm{F555W-F814W})^\mathrm{WFPC2} = 0.0129x^2 + 0.6068x + 0.4000 \label{eq:wfpc2-decals}
\end{equation}

where $x = (g-z)^\mathrm{DECam}$. These relations have $\sigma_\mathrm{(F475W-F814W)^{ACS}} = 0.03$, $\sigma_{(g-i)^{\mathrm{CFHT}}} = 0.04$, and $\sigma_\mathrm{(F555W-F814W)^{WFPC2}} = 0.01$ for $(g-z)^\mathrm{DECam} \leq 1.5$ and $\sigma_\mathrm{(F475W-F814W)^{ACS}} = 0.02$, $\sigma_{(g-i)^{\mathrm{CFHT}}} = 0.03$, and $\sigma_\mathrm{(F555W-F814W)^{WFPC2}} = 0.01$ for $(g-z)^\mathrm{DECam} \leq 1$, noting that there are fewer blue SSPs against which to compare, so the larger scatter among redder stellar populations is likely driven by the wider range of individual age and metallicity combinations that result in those photometric colors.

\section{Ion-to-ion column densities in the dwarf CGM}\label{app:ions}

\begin{figure*}
    \centering
    \includegraphics[width = \linewidth]{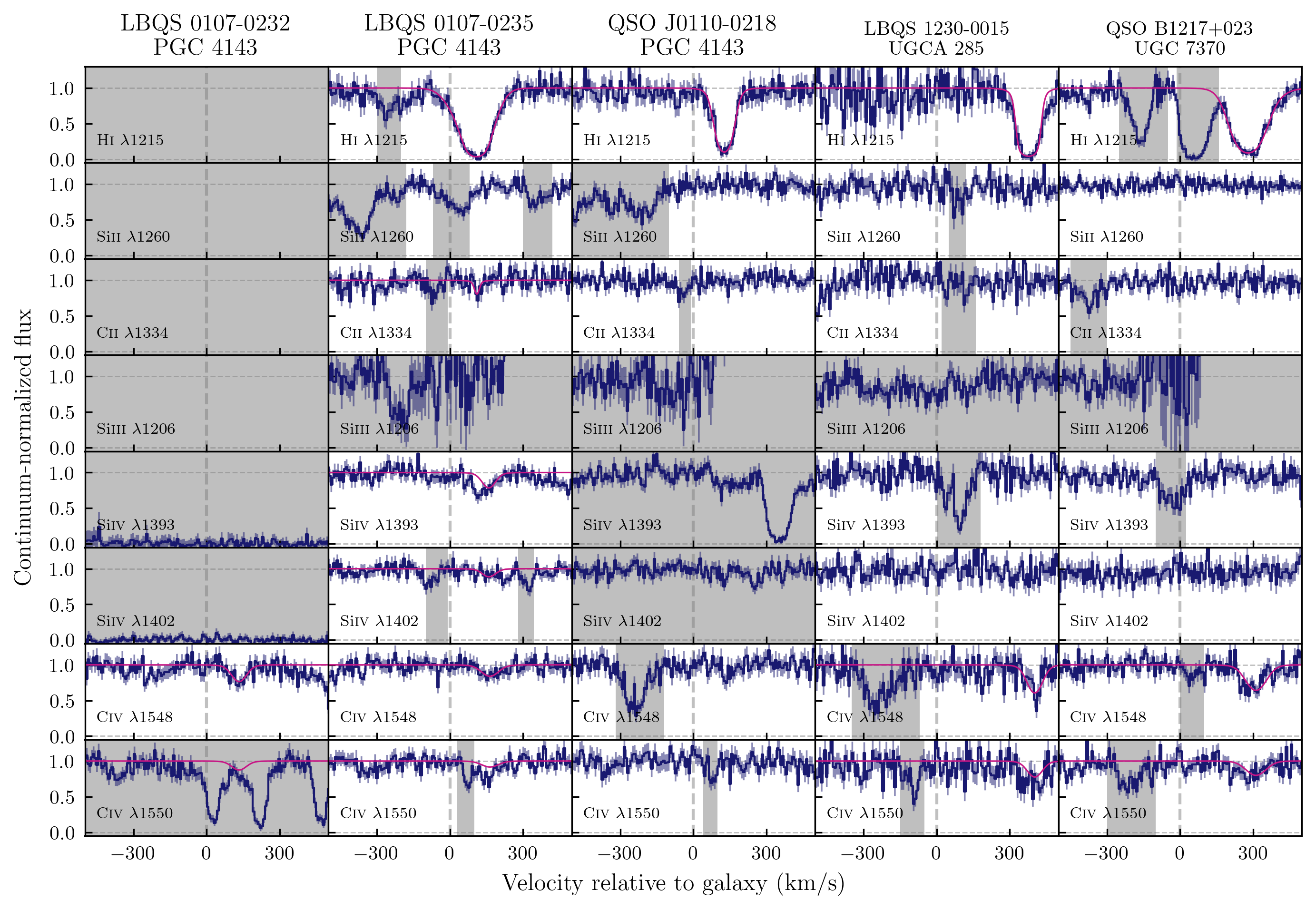}
    \caption{Same as Figure \ref{fig:dwarf_cgm} for the sightlines associated with PGC 4143, UGCA 285, and UGC 7370. Note that complete non-detections where it is not possible to place a limit on the line strength (due, in general, to wavelength coverage, the Lyman limit, or significant contamination) are denoted by a grayed out panel.
    \label{fig:abs1}}
\end{figure*}

\begin{figure*}
    \centering
    \includegraphics[width = \linewidth]{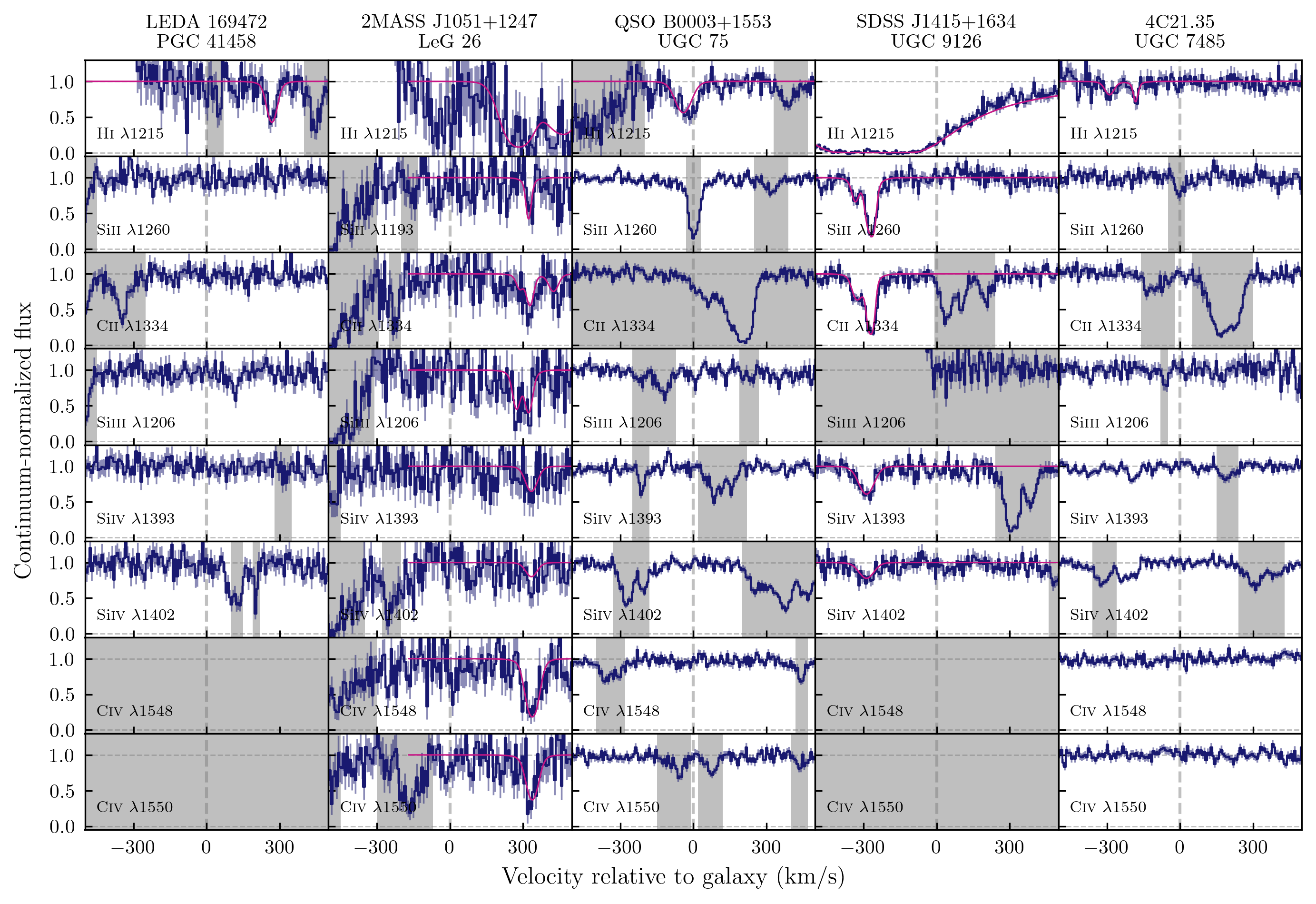}
    \caption{Same as Figure \ref{fig:dwarf_cgm} for the sightlines associated with PGC 41458, LeG 26, UGC 75, UGC 9126, and UGC 7485. Note that the \ion{Si}{2} measurement for LeG 26 comes from \ion{Si}{2} $\lambda 1193$ instead of \ion{Si}{2} $\lambda 1260$.
    \label{fig:abs2}}
\end{figure*}

\begin{figure*}
    \centering
    \includegraphics[width = \linewidth]{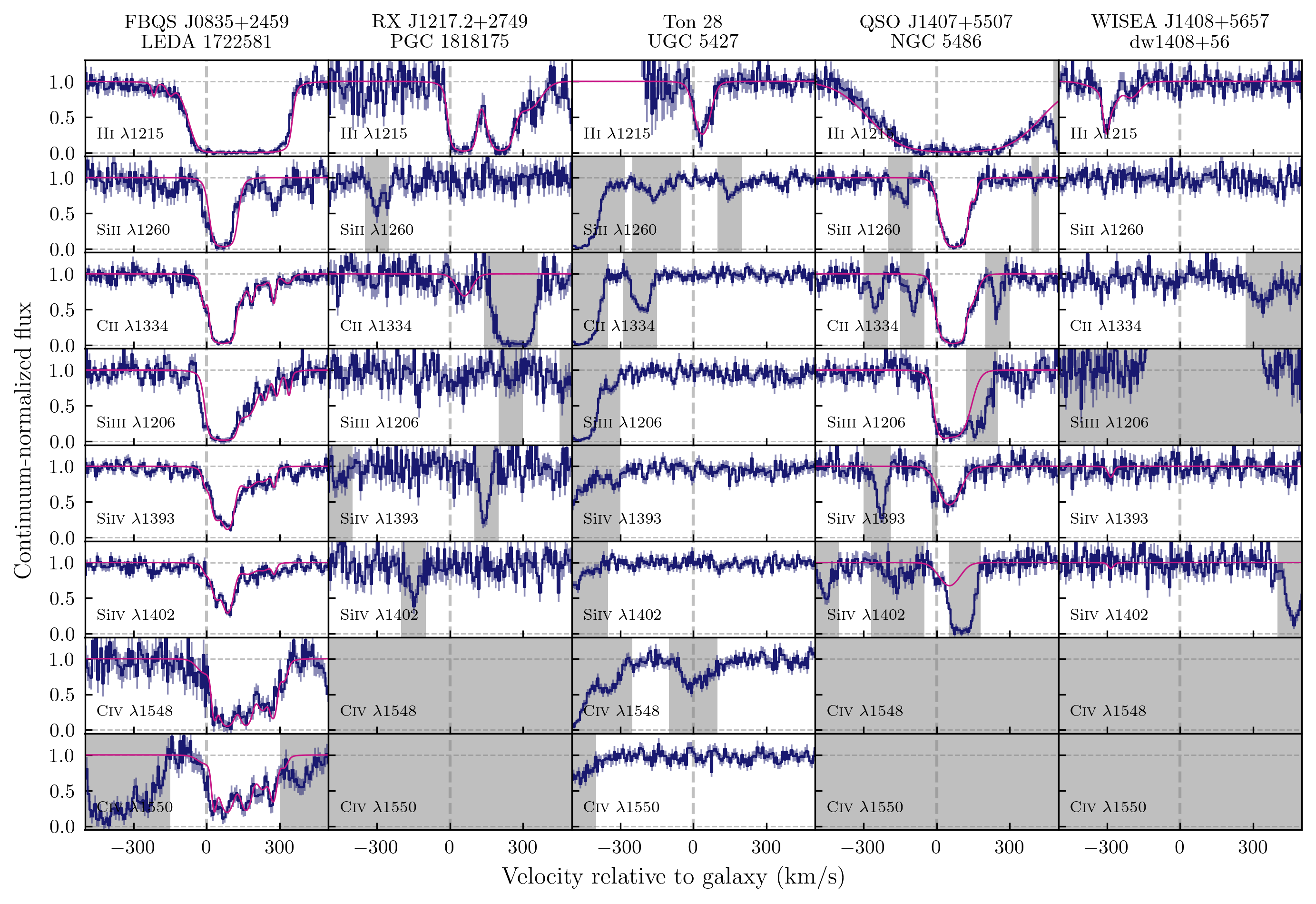}
    \caption{Same as Figure \ref{fig:dwarf_cgm} for the sightlines associated with LEDA 1722581, PGC 1818175, UGC 5427, NGC 5486, and dw1408+56.
    \label{fig:abs3}}
\end{figure*}

\begin{figure*}
    \centering
    \includegraphics[width = \linewidth]{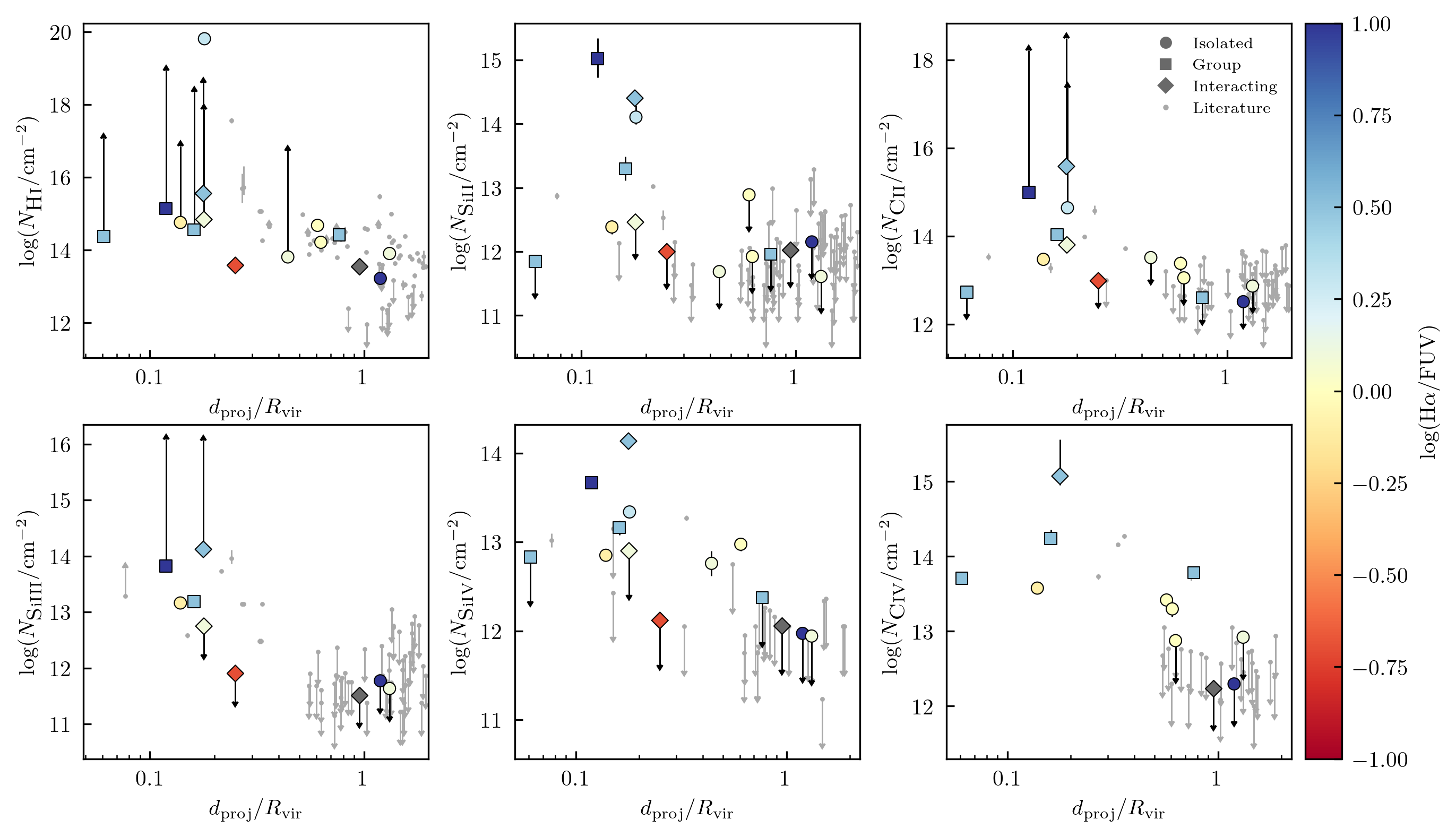}
    \caption{Same as Figure \ref{fig:cgm_mass}, but with QUEST Dwarfs galaxies colored by their $\log(\mathrm{H}\alpha/\mathrm{FUV})$.
    \label{fig:cgm}}
\end{figure*}

\begin{figure*}
    \centering
    \includegraphics[width = \linewidth]{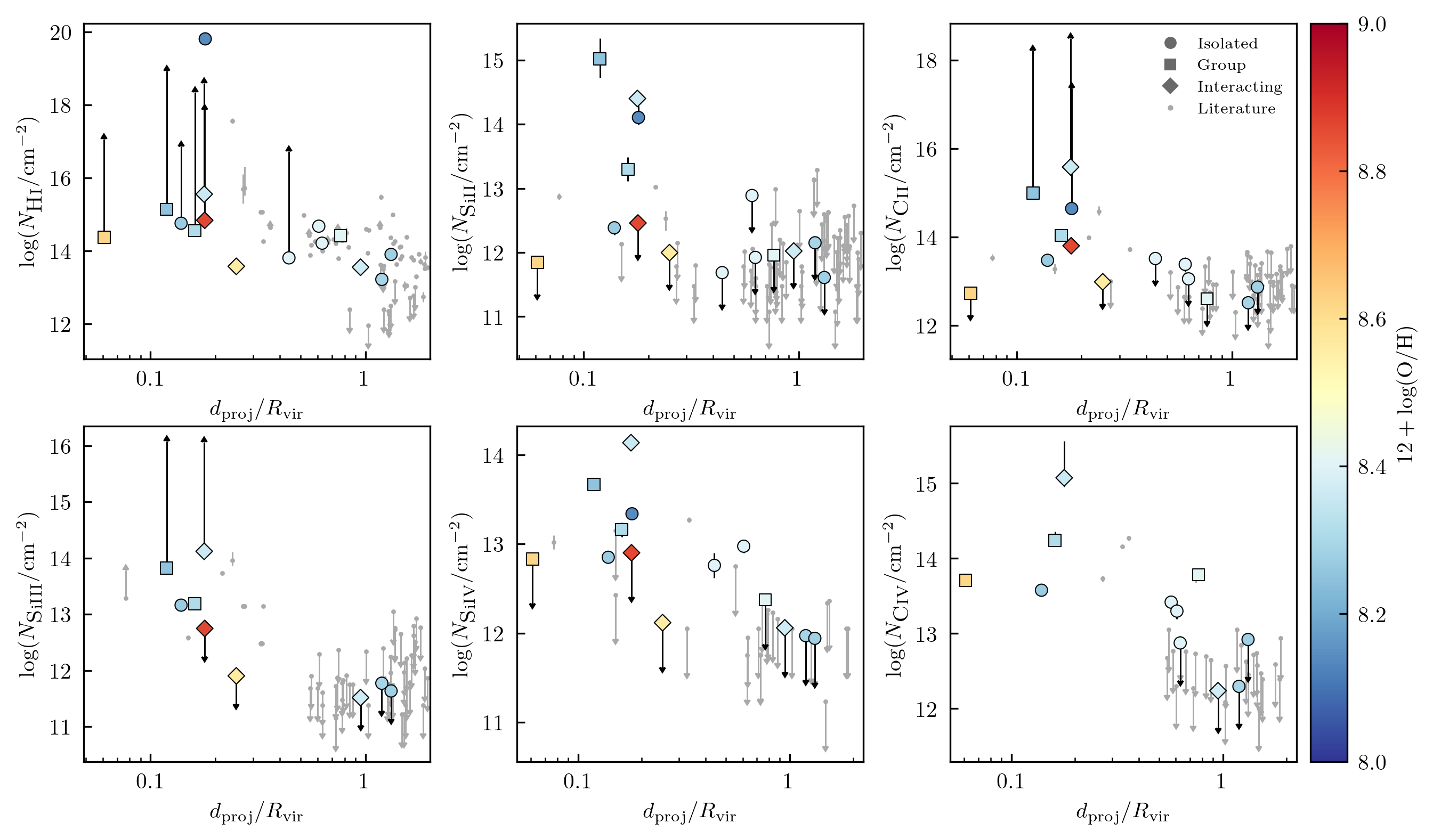}
    \caption{Same as Figure \ref{fig:cgm_mass}, but with QUEST Dwarfs galaxies colored by $12+\log(\mathrm{O/H})$.
    \label{fig:cgm_met}}
\end{figure*}

Individual ion column density measurements (detailed in Table \ref{tab:ion}), color-coded based on \mstar\ have been shown in Figure \ref{fig:cgm_mass}.  Here we present the underlying FUV absorption spectra (Figures \ref{fig:abs1}, \ref{fig:abs2}, and \ref{fig:abs3}) and the ionic column density density profiles with the measurements color-coded by $\log(\mathrm{H}\alpha/\mathrm{FUV})$ (Figure \ref{fig:cgm}) and $12 + \log(\mathrm{O/H})$ (Figure \ref{fig:cgm_met}). 

\bibliography{references}{}
\bibliographystyle{aasjournal}

\end{document}